\documentclass[10pt,twocolumn,amsmath,amssymb,aps,pra,showkeys,superscriptaddress,floatfix]{revtex4-2}
\usepackage{epsfig}
\usepackage{color}
\usepackage{amsmath}
\usepackage[colorlinks,linkcolor=blue,citecolor=blue,urlcolor=blue]{hyperref}
\renewcommand{\vec}[1]{\underline{#1}}

\usepackage[english]{babel}
\usepackage{float}

\usepackage{xr-hyper}
\begin{document}

\title{Machine learning model for efficient nonthermal tuning of the charge density wave in monolayer NbSe$_2$}

\author{Luka Benić}
\affiliation{Ruđer Bošković Institute, Zagreb, Croatia}

\author{Federico Grasselli}
\affiliation{Dipartimento di Scienze Fisiche, Informatiche e Matematiche, Universit\`a degli Studi di Modena e Reggio Emilia, Modena, Italy}
\affiliation{CNR NANO S3, Modena, Italy}

\author{Chiheb Ben Mahmoud}
\affiliation{Inorganic Chemistry Laboratory, Department of Chemistry, University of Oxford, Oxford OX1 3QR, United Kingdom}

\author{Dino Novko}
\email{dino.novko@gmail.com}
\affiliation{Centre for Advanced Laser Techniques, Institute of Physics, Zagreb, Croatia}

\author{Ivor Lončarić}
\email{ivor.loncaric@gmail.com}
\affiliation{Ruđer Bošković Institute, Zagreb, Croatia}

\begin{abstract}
Understanding and controlling the charge density wave (CDW) phase diagram of transition metal dichalcogenides is a long-studied problem in condensed matter physics. However, due to complex involvement of electron and lattice degrees of freedom and pronounced anharmonicity, theoretical simulations of the CDW phase diagram at the density-functional-theory level are often numerically demanding. To reduce the computational cost of first principles modelling by orders of magnitude, we have developed an electronic free energy machine learning model for monolayer NbSe$_2$ that allows changing both electronic and ionic temperatures independently. Our approach relies on a machine learning model of the electronic density of states and zero-temperature interatomic potential. This allows us to explore the CDW phase diagram of monolayer NbSe$_2$ both under thermal and laser-induced nonthermal conditions. Our study provides an accurate estimate of the CDW transition temperature at low cost and can disentangle the role of hot electrons and phonons in nonthermal ultrafast melting process of the CDW phase in NbSe$_2$.

\end{abstract}  

\maketitle

\section{Introduction}
\label{sec:introduction}
In addition to having exceptional optical properties, transition-metal dichalcogenides (TMDs) have gained a lot of attention due to their complex phase diagram, containing various charge-ordered phases, such as charge density waves (CDW), superconductivity, and even signs of order fluctuations\,\cite{rossnagel2011origin,manzeli17tmds}. One notable example is NbSe$_2$, which supports both CDW and superconductivity already at ambient pressure\,\cite{foner1973sc,harper1975cdw}, while both can be modified or quenched with an increase in applied pressure or temperature\,\cite{Cho2018scdw,moulding2020}. 
An interesting interplay between CDW and superconductivity was observed when the number of layers in NbSe$_2$ is varied from bulk up to the monolayer limit\,\cite{staley2009sc,Xi2015cdw}. For instance, the value for superconductivity temperature $T_{\rm SC}$ drops from 7.2\,K in bulk, to 3K in the monolayer\,\cite{staley2009sc}. The corresponding modifications to the CDW phase when NbSe$_2$ is thinned down to a monolayer limit are not so clear, and are still a matter of active debate\,\cite{Xi2015cdw,Ugeda2015cdw,Lin2020cdw,azoury2023}. Namely, while certain experimental measurements suggest that the CDW transition temperature $T_{\rm CDW}$ in the NbSe$_2$ single layer is $T_{\rm CDW}=145$\,K, namely significantly increased compared to the bulk\,\cite{Xi2015cdw,Lin2020cdw,azoury2023}, there are also reported experimental evidences that the CDW phase remains intact and $T_{\rm CDW}$ is close to the value of 33.5\,K as in the bulk\,\cite{Ugeda2015cdw}. In order to resolve this controversy, a thorough quanatitive and efficient computational study needs to be performed that is able to incorporate all the relevant many-body interactions ruling the CDW phase in NbSe$_2$. 

Several theoretical studies have inspected the origin of the CDW in NbSe$_2$ and have clearly proven that there are no electronic instabilities in NbSe$_2$ that would lead to the standard Peierls transition, while the CDW order is induced mainly by the orbital-dependent and anisotropic electron-phonon coupling\,\cite{johannes2006,xuetao2015,flicker2015charge}, as also confirmed by the experiments\,\cite{valla2004,weber2011origin}. 
On top of that, anharmonic effects were shown to be important for the proper and quantitative description of the CDW phase transition\,\cite{mcmillan1977,varma1983,flicker2015charge}. Furthermore, \emph{ab initio} anharmonic study based on stochastic self-consistent harmonic approximation (SSCHA), which includes quantum and thermal fluctuations at a non-perturbative level, reported weak dimensionality effect in NbSe$_2$, with $T_{\rm CDW} = 59$\,K and 73\,K for bulk and monolayer, respectively\,\cite{leroux2015,PhysRevLett.125.106101}. On the other hand, the results based on \emph{ab-initio} and path-integral molecular dynamics showed more stronger dimensionality dependence with $T_{\rm CDW}=25-50$\,K for bulk, and $T_{\rm CDW}=75-100$\,K for the monolayer\,\cite{zheng22}. With this, it is obvious that the dimensionality controversy in NbSe$_2$ is still not settled. Even though both of these theoretical approaches are suitable to resolve this issue, they are computationally demanding and it is not easy to reach the numerical convergence.

A powerful approach to control the CDW state, as well as to unravel the origin of CDW transition is laser-induced ultrafast dynamics, where it is possible to separate electron and phonon contributions to the CDW in time and with different excitation conditions\,\cite{hellmann2012time,porer14,otto21}. 
Commonly, an ultrafast sub-picosecond rise or decay of the signal in time-resolved photoemission and transient optical spectroscopy is attributed to the purely electron-related origin of the CDW\,\cite{monney16}, such as excitonic insulator\,\cite{jerome67}, while slower picosecond dynamics to the phonon-driven mechanisms, namely electron-phonon and phonon-phonon interactions\,\cite{hellmann2012time}. Recent ultrafast time-resolved measurements in NbSe$_2$ reported a fast transient electronic component, the existence of the CDW amplitude phonon mode, melting excitation density, involvement of strong electron-phonon coupling, and slow dynamics of order parameter\,\cite{anikin2020,payne2020,watanabe2022,venturini2023}. These experimental observations suggest an unconventional involvement of various many-body interactions in photo-excited NbSe$_2$, however, so far there are no \emph{ab initio} theoretical studies that could provide a deeper insight and tackle electron- and lattice-related fluctuation in non-thermal conditions.

Modelling phase diagrams of materials efficiently and from first principles is a long-sought-after goal of computational material science. Machine learning (ML) interatomic potentials have recently reached the same accuracy as the baseline density functional theory (DFT) on which they are trained to predict total energies and forces, while offering orders of magnitude faster evaluation. Given that the training dataset samples the configurational space well, one could, in principle, use the resulting potential to efficiently model the phase diagram of any material across pressures and temperatures. This has been successfully demonstrated for a few materials or classes of materials such as elemental systems \cite{zong2018developing}, alloys \cite{rosenbrock2021machine}, high-entropy alloys \cite{kostiuchenko2019impact}, hybrid inorganic-organic systems \cite{braeckevelt2022accurately}, and molecular crystals \cite{mladineo2024thermosalient}.
However, important features of the phase diagram of some materials, such as CDW ordering, have not yet been modelled from the first principles with ML interatomic potentials. 

One of the limitations in the current machine learning interatomic potentials is that they are trained on DFT data for fixed electronic temperature $T_{el}$. This $T_{el}$ is conveniently used to improve the convergence of DFT calculation and different smearing methods exist, from Fermi-Dirac smearing representing the real, but in practice very large $T_{el}$ up to methods that try to better represent $T_{el}\approx0$ K case.
However, $T_{el}$ has an important effect on the potential energy surface and, thus, the phase diagram of NbSe$_2$ and similar materials. One solution would be to train a separate ML for each electronic temperature set by the Fermi-Dirac distribution. This is highly impractical for modelling phase diagram as a function of temperature. Additionally, setting a low Fermi-Dirac smearing temperature requires dense sampling of the Brillouin zone, making DFT calculations computationally more demanding.
Moreover, if one would study the phase diagram under ultrafast laser conditions, ion and electron temperatures have to be separately controlled as light couples differently to electron and ion subsystems.

To solve these general issues, and to resolve some of the long-standing questions in the NbSe$_2$ phase diagram, we have constructed an ML model for electronic free energy that enables independent control of both electron and phonon temperatures. Our ML model is based on learning the $T_{el}=0$ K ML interatomic potential from DFT energies and forces and learning the ML model of the electronic density of states at $T_{el}=0$ K \cite{PhysRevB.106.L121116}. From these two ML models it is then possible to calculate the electronic free energy potential for any $T_{el}$. We have used our model to compute the phase diagram of NbSe$_2$ as a function of electronic and ionic temperatures that compares with previous DFT-based calculations at orders of magnitude lower computational cost. In particular, for equilibrium conditions we get $T_{\rm CDW}\approx 104$\,K, which agrees with \emph{ab initio} molecular dynamics results\,\cite{zheng22} that suggest a finite dimensionality effect in NbSe$_2$.
We have then simulated response of NbSe$_2$ to a short laser pulse within the two-temperature model (TTM) based on DFT. Our model reveals a strong exchange of energy between lattice and electrons, which leads to sub-picosecond dynamics of electron component, and long-lived phonon part. In combination with the two-temperature ML interatomic potential, we have shown that hot electrons can efficiently melt the electron-phonon-driven CDW order in NbSe$_2$. These results introduce a fresh perspective on the understanding of time-resolved measurements, showing that it is not necessary to invoke additional ordering mechanisms of purely electronic origin to explain the sub-picosecond decay of the CDW-related signals.

\section{Models and Methods}
\label{sec:models_and_methods}

\subsection{Theoretical model}
\label{subsec:theoretical_model}
To efficiently model finite electronic temperature effects, we use an approximation to the exact Helmholtz electronic free energy that depends only on quantities at $T_{el}=0$ K. As shown in the Supplementary Material and in Ref. \cite{PhysRevB.106.L121116}, the Helmholtz electronic free energy of the system with $N_{ion}$ ions and $N_{el}$ electrons can be approximated by
\begin{align}
    & F(\vec{\boldsymbol{R}}; T_{el}) \approx E^{0}(\vec{\boldsymbol{R}}; 0) + \Delta F(\vec{\boldsymbol{R}};0,T_{el})~.\label{eq:energy}
\end{align}
where $\vec{\boldsymbol{R}} = \{\boldsymbol{R}_{1},...,\boldsymbol{R}_{N_{ion}}\}$ represents the set of ionic coordinates of a given system (we will call $\vec{\boldsymbol{R}}$ a structure from now on), $E^{0}(\vec{\boldsymbol{R}}; 0)$ is the $T_{el}=0$ K DFT energy in its standard form \cite{giustino2014materials} and $\Delta F(\vec{\boldsymbol{R}}; 0, T_{el})$ is the finite $T_{el}$ correction to the Helmholtz electronic free energy, constructed out of the ground state electronic density of states (eDOS), which is defined as
\begin{align}
    g^{0}(\vec{\boldsymbol{R}}; \epsilon) = \sum_{s,n, \boldsymbol{k}}\delta(\epsilon-\epsilon^{0}_{s, n, \boldsymbol{k}}(\vec{\boldsymbol{R}}))~,\label{eq:edos}
\end{align}
where the sum runs over the spin projections $s$, bands $n$, and k-points $\boldsymbol{k}$. The force $\boldsymbol{F}_{n}(\vec{\boldsymbol{R}}; T_{el})$ on the $n$-th ion, represented by the coordinates $\boldsymbol{R}_{n}$, is given by the negative gradient with respect to $\boldsymbol{R}_{n}$ of \eqref{eq:energy}
\begin{align}
    \boldsymbol{F}_{n}(\vec{\boldsymbol{R}}; T_{el})=-\nabla_{\boldsymbol{R}_{n}}F(\vec{\boldsymbol{R}}; T_{el})~\label{eq:force}
\end{align}
and is given in the Supplementary Material.

The Helmholtz electronic free energy $F(\vec{\boldsymbol{R}}; T_{el})$ in this approximation is fully determined by the $T_{el}=0$ K ground state energy $E^{0}(\vec{\boldsymbol{R}}; 0)$ and the finite electronic temperature correction ($\Delta F(\vec{\boldsymbol{R}};0,T_{el})$) constructed from the ground state eDOS $g^{0}(\vec{\boldsymbol{R}}; \epsilon)$. Furthermore, the ionic force $\boldsymbol{F}_{n}(\vec{\boldsymbol{R}}; T_{el})$ is determined by the gradients of $E^{0}(\vec{\boldsymbol{R}}; 0)$ and the gradients of $g^{0}(\vec{\boldsymbol{R}}; \epsilon)$.

The given approximation opens up the possibility to construct an electronic free energy ML model using the $T_{el}=0$ K (ground state) DFT total energies (and forces) and DFT electronic density of states that would be able to predict the behaviour of a material at any electronic temperature.

\subsection{Training dataset}
\label{subsec:training_dataset}

To construct the dataset we performed DFT calculations using the Vienna Ab initio Simulation Package (VASP) software version 6.4.2 \cite{PhysRevB.47.558, KRESSE199615, PhysRevB.54.11169, G_Kresse_1994, PhysRevB.59.1758} with the Perdew-Burke-Ernzerhof (PBE) functional \cite{PhysRevLett.77.3865}. For the plane-wave basis set energy cutoff we used 270 eV which is 1.3 times larger than the default value for the VASP Se and Nb\_pv pseudopotentials. 
To model $T_{el}=0$ K, we have used the Methfessel-Paxton scheme of order $1$ with a smearing of 0.0043 eV.
To obtain accurate results, we used fine sampling of the Brillouin zone of $93\times93\times1$ for the unit cell, and for the larger supercells it was scaled accordingly.
To collect data, we performed molecular dynamics with the Andersen thermostat \cite{10.1063/1.439486} spanning temperatures from 10 to 300 K and the Bayesian-learning algorithm for on-the-fly machine learning as implemented in VASP.

The dataset consists of $2452$ different structures in total, with supercell sizes of $2\times2\times1$, $3\times3\times1$, $4\times4\times1$, $6\times6\times1$, $9\times9\times1$, $6\times1\times1$, $1\times9\times1$, $12\times1\times1$ and $2\times9\times1$. The dataset was randomly split into $80\%$ training and $20\%$ test sets.
Lattice constant for calculations was 3.4715 \AA. A small fraction of $3\times3\times1$ structures in the dataset were strained, with uniaxial strains in the range of negative $1.8\%$ to positive $1.2\%$.

\subsection{Machine learning interatomic potential at $T_{el}=0$ \textnormal{K}}
\label{subsec:E0_0K}

The model for $E^{0}(\vec{\boldsymbol{R}}; 0)$ and its gradients was constructed with the message-passing atomic cluster expansion (MACE) architecture \cite{Batatia2022Design, NEURIPS2022_4a36c3c5}. Most of the training hyperparameters were kept to the default values (see Appendix A of \cite{NEURIPS2022_4a36c3c5}).
To be able to accurately model acoustic phonons, the cutoff radius was set to 8~\AA~and we used 8 Bessel basis functions, 128 embedding channels and $l_{max}=2$, namely, we included equivariant messages up to the second order. In the weighted loss function, energy ($\lambda_{E}$) and forces ($\lambda_{F}$) weights were set to 1 and 100, respectively. The model was trained for 2000 epochs, with batch size of 5. 80\% of the original structures were taken as a training set, with 5\% of them taken for the validation set, and the other 20\% of the original structures comprised a test set. Stochastic weight averaging \cite{izmailov2019averagingweightsleadswider, athiwaratkun2019consistentexplanationsunlabeleddata} was performed for the last 20\% of epochs, and in this part of training $\lambda_{E}$ weight was set to 1000 and $\lambda_{F}$ weight was kept at 100.

\begin{figure}[h]
\centering
\includegraphics[width=\columnwidth]{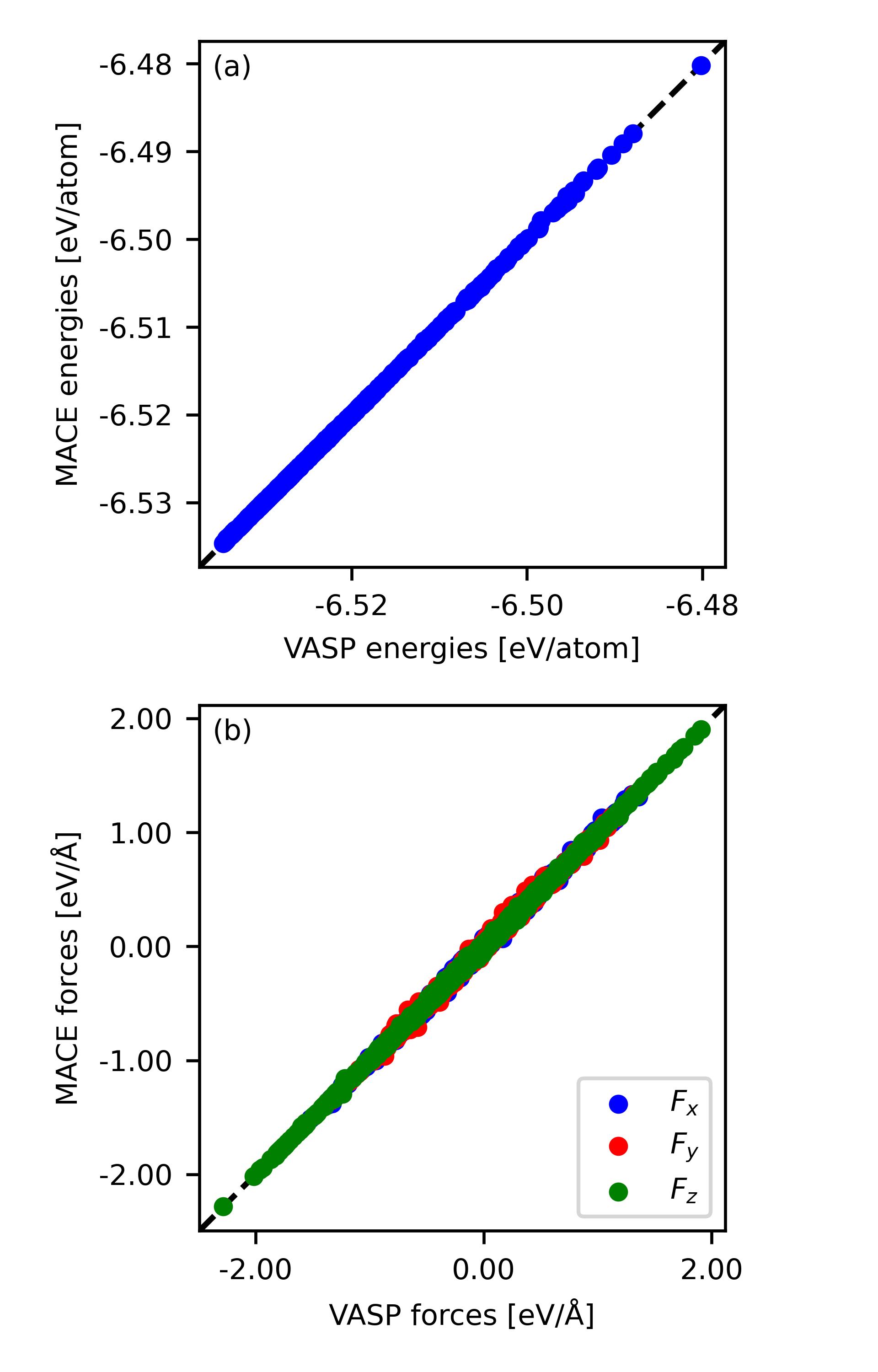}
\caption{(a) A correlation plot between the VASP and MACE energies evaluated on the test set. (b) A correlation plot between the VASP and MACE forces per component evaluated on the test set.}
\label{fig:mace_subplots}
\end{figure}
In Fig. \ref{fig:mace_subplots} we show the correspondence between the MACE and the VASP results for $E^{0}(\vec{\boldsymbol{R}}; 0)$ evaluated on the test set. The performance metrics evaluated on the test set are shown in Table \ref{table:mace}.
\begin{table}[H]
\caption{The performance of the MACE model evaluated on the test set.}
\label{table:mace}
\centering
\begin{tabular}{|c|c|c|c|}
\hline
\text{quantity} & \text{MAE} & \text{RMSE} & \text{R}$^{2}$ \\
\hline\hline
\text{energy}&$0.04$ \text{meV/atom}&$0.06$ \text{meV/atom}&$1.00$\\
\hline
\text{F$_{x}$}&$4.10$ \text{meV/Å}&$7.07$ \text{meV/Å}&$0.99$\\
\hline
\text{F$_{y}$}&$4.08$ \text{meV/Å}&$7.51$ \text{meV/Å}&$0.99$\\
\hline
\text{F$_{z}$}&$2.79$ \text{meV/Å}&$4.76$ \text{meV/Å}&$1.00$\\
\hline
\end{tabular}
\end{table}
The accuracy of $<$0.1 meV/atom and $<$10 meV/\AA~is considered state-of-the-art \cite{10.1063/5.0155322}. 
However, CDW instability is related to acoustic phonons, which are typically much harder to converge.
\begin{figure}[h]
\center
\includegraphics[width=\columnwidth]{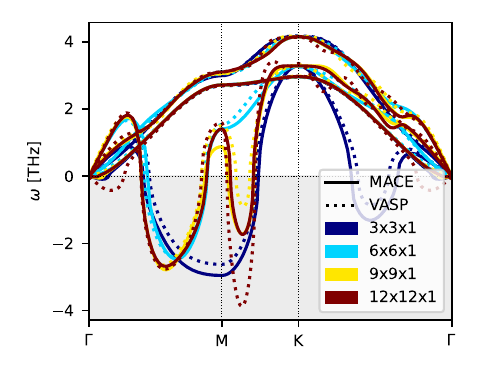}
\caption{Comparison of the acoustic phonon dispersions for different supercell sizes at $T_{el}=0$ K in the harmonic approximation between the MACE and VASP.}
\label{fig:phonons_cells}
\end{figure}
We calculated finite-difference DFT phonon dispersions in the harmonic approximation for different supercell sizes, as shown in Fig. \ref{fig:phonons_cells}. We used the Phonopy code \cite{Togo_2023} with a finite difference step of 0.015 \AA.
Clearly, to converge the complete phonon dispersion, one would have to use supercells larger than $12\times12\times1$ making DFT calculations expensive.
This points to long-range interactions in NbSe{$_2$}, arising from its two-dimensional nature, namely weak screening. As discussed below in detail, the important instability that gives rise to the CDW order between the $\Gamma$ and M points is already converged for the $9\times9\times1$ supercell. The instability between M and K points is quickly removed with anharmonic corrections and, therefore, is not of particular importance. Previous DFT-based works also used supercells up to $9\times9\times1$ \cite{PhysRevLett.125.106101} and therefore we aim to reproduce these results with our ML potential.

The fact that interactions up to 9 unit cells ($>30$ \AA) are important for converging acoustic phonons makes the construction of machine learning interatomic potential challenging, since typically a cutoff radius of 5-10~\AA~ has to be imposed. Actually, we are not aware of any trained potential for this class of materials despite both the importance of these materials and large increase of trained potentials in recent years.
Using the message-passing and the relatively large initial cutoff of 8~\AA~as well as including the data of large supercells, we have managed to converge our machine learning potential to reproduce DFT phonons, as shown in Fig. \ref{fig:phonons_cells}.

\subsection{Machine learning electronic density of states at $T_{el}=0$ \textnormal{K}}
\label{subsec:eDOS_0K}

The model for eDOS $g^{0}(\vec{\boldsymbol{R}}; \epsilon)$ and its gradients was trained using a Kernel Ridge Regression (KRR) model \cite{acs.chemrev.1c00022} as implemented in \textit{librascal} \cite{Musil2021EfficientIO}, and Smooth Overlap of Atomic Positions (SOAP) descriptors \cite{PhysRevB.87.184115}. 
\begin{figure*}[ht]
\centering
\includegraphics[width=2\columnwidth]{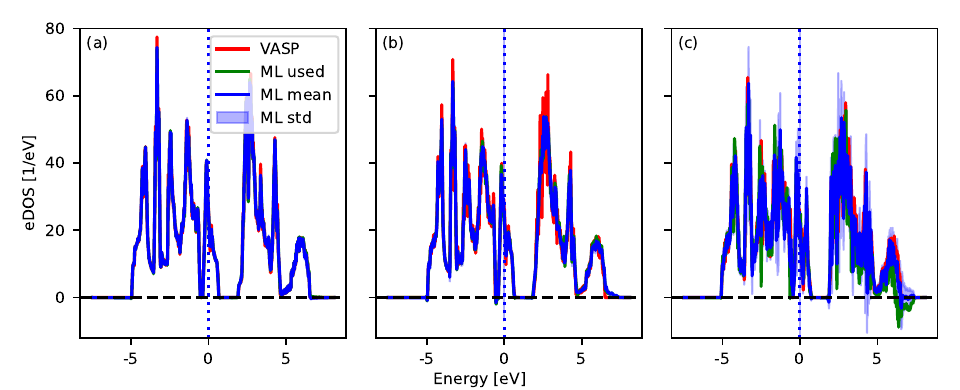}
\caption{Results of the $T_{el}=0$ K eDOS predictions for an ensemble of 10 models with the smallest errors \eqref{eq:rmse} on the validation set during the Bayesian hyperparameter optimization process using Optuna  \cite{akiba2019optunanextgenerationhyperparameteroptimization}. From left to right: (a) the best, (b) median and (c) the worst predictions (with respect to the RMSE) on the test set. Red lines represent the baseline eDOS constructed from the VASP data, green lines represent the eDOS predicted by the model chosen (\%RMSE on the validation set of this model is 14.49\%) for the calculations done in \ref{sec:results_and_discussions}, blue lines represent the means of 10 models and light blue represents the standard deviation of the 10 model predictions. Vertical dotted lines represent the Fermi energies. One can observe that the largest deviations between the ML and DFT eDOS occur well beyond the Fermi level and are likely to impact calculations only at extremely high temperatures.}
\label{fig:edos}
\end{figure*}
As in Ref. \cite{PhysRevB.102.235130} the ML model for the eDOS leverages a decomposition into local contributions from each ionic environment specified by its ionic coordinates $\boldsymbol{R}_{i}$
\begin{equation}
    g^{0}(\vec{\boldsymbol{R}}; \epsilon) = \sum_{\boldsymbol{R}_{i}\in \vec{\boldsymbol{R}}} g^{0}(\boldsymbol{R}_{i};\epsilon)\stackrel{\eqref{eq:edos}}{=}\sum_{\substack{s,n, \boldsymbol{k} \\ \boldsymbol{R}_{i}\in \vec{\boldsymbol{R}}}}\delta(\epsilon-\epsilon^{0}_{s, n, \boldsymbol{k}}(\boldsymbol{R}_{i}))~.\label{eq:local_edos}
\end{equation}
The Dirac delta distributions in \eqref{eq:local_edos} are approximated by the Gaussian distributions, defined by the Gaussian broadening hyperparameter $g_{b}$.
\begin{equation}
    \delta(\epsilon-\epsilon^{0}_{s, n, \boldsymbol{k}}(\boldsymbol{R}_{i})) \rightarrow \frac{1}{\sqrt{2\pi g^{2}_{b}}}e^{-\frac{(\epsilon - \epsilon^{0}_{s, n, \boldsymbol{k}}(\boldsymbol{R}_{i}))^{2}}{2 g^{2}_{b}}}~.\label{eq:gaussian}
\end{equation}
In this work, the value of $g_{b}=0.01$ eV was chosen. For the construction of eDOS, we use a hyperparameter $d\epsilon = 0.00025$ eV to control the discretization of the energy grid. These two hyperparameters were set to low enough values to ensure a detailed eDOS and a finely discretized energy grid, whenever precise numerical integration is needed, see the Supplementary Material. In Fig. S2 and Fig. S3 comparison between the approximation and the VASP results for the Helmholtz electronic free energy for different values of $g_{b}$ and $d\epsilon$ with $T_{el}=500$ K are reported. As can be seen, the values we have chosen for this work are the most suitable. Both of these hyperparameters should be part of the hyperparameter optimization process, but that would be computationally too expensive.

Since all the terms in \eqref{eq:energy} and \eqref{eq:force} are invariant to the choice of zero energy in eDOS, there is freedom in the choice of zero energy in the eDOS construction. We have chosen to align the eDOS of every structure to its Fermi energy, namely the Fermi energy is at $\epsilon=0$ eV for every structure in the dataset. In the literature, there are other options to align the eDOS, two examples are alignment to the core electron eDOS \cite{PhysRevB.106.L121116} and the adaptive energy reference \cite{PhysRevMaterials.9.013802}. We also performed our calculations using alignment of the eDOS to the core electron eDOS, but there were no noticeable differences in the results.
Furthermore, the cutoff radius for the SOAP descriptors for this ML model was set to 7.6 \AA~after the Bayesian hyperparameter optimization process. Other relevant hyperparameters are given in the supplementary note S2.

Due to the high memory consumption of the full KRR models, sparse KRR was used during the training of an ML model for eDOS. The sparsification of ionic environments to construct the so-called active set ($\vec{\boldsymbol{A}}$) was performed using the Farthest Point Sampling algorithm \cite{623193} and the number of sparse ionic environments was also part of the Bayesian hyperparameter optimization process. There are 545 Se and 272 Nb ionic environments in the active set. The eDOS of the $i$-th local ionic environment $\boldsymbol{R}_{i}$ of a structure $\vec{\boldsymbol{R}}$ can be expanded on a set of positive-definite functions (kernel functions), using the set $\vec{\boldsymbol{A}}$ \cite{item_d4185ca57bd14f11989d9b59bb61c5d2},
\begin{equation}
    g^{0}(\boldsymbol{R}_{i}; \epsilon) = \sum_{\boldsymbol{A}_{j}\in \vec{\boldsymbol{A}}} x_{j}(\epsilon)k(\boldsymbol{A}_{j}, \boldsymbol{R}_{i})~,\label{eq:local_kernel_edos}
\end{equation}
where with $\boldsymbol{A}_{j}\in\vec{\boldsymbol{A}}$ we note the \textit{j}-th ionic environment from the active set, $x_{j}(\epsilon)$ are the weights of the model and $k(\boldsymbol{A}_{j}, \boldsymbol{R}_{i})$ are the kernel functions. 

The kernel functions are defined using the SOAP framework and are given in \cite{PhysRevB.87.184115, PhysRevB.102.235130}. SOAP power spectrum was raised to a power of $\zeta > 1$, to effectively include higher-body order correlations, and $\zeta$ was part of the Bayesian hyperparameter optmization process. In this work, the value of $\zeta=2$ was used.

From the first equality in \eqref{eq:local_edos} and \eqref{eq:local_kernel_edos} we have
\begin{equation}
    g^{0}(\vec{\boldsymbol{R}}; \epsilon) = \sum_{\substack{\boldsymbol{R}_{i}\in \vec{\boldsymbol{R}} \\ \boldsymbol{A}_{j}\in \vec{\boldsymbol{A}}}} x_{j}(\epsilon)k(\boldsymbol{A}_{j}, \boldsymbol{R}_{i})=\boldsymbol{k}^{T}_{\vec{\boldsymbol{R}}\hspace{0.5mm}\vec{\boldsymbol{A}}}\cdot \boldsymbol{x}_{\vec{\boldsymbol{A}}}(\epsilon)~.\label{eq:kernel_edos}
\end{equation}
From \eqref{eq:kernel_edos} we can conclude \cite{PhysRevB.102.235130}
\begin{equation}
    \nabla_{\boldsymbol{R}_{i}}g^{0}(\vec{\boldsymbol{R}}; \epsilon) = (\nabla_{\boldsymbol{R}_{i}}\boldsymbol{k}^{T}_{\vec{\boldsymbol{R}}\hspace{0.5mm}\vec{\boldsymbol{A}}})\cdot \boldsymbol{x}_{\vec{\boldsymbol{A}}}(\epsilon)~.\label{eq:gradient_edos}
\end{equation}
From \eqref{eq:gradient_edos} we can see that in this model it is easy to obtain the gradients of the eDOS with respect to the ionic positions, and thus free energy forces, which is a great advantage of this model. In our dataset we did not include any eDOS gradients, even though it could increase the precision of the model. This is because the gradients of the eDOS with DFT can be obtained only by finite differences, making it expensive and inconvenient.

To evaluate the performance of the model during training and optimize the hyperparameters, we use the metric proposed in \cite{PhysRevB.102.235130}
\begin{equation}
    \%\text{RMSE} = \frac{\sqrt{\frac{1}{N}\sum^{N}_{i}\int(\widetilde{g^{0}}(\vec{\boldsymbol{R}}_{i}; \epsilon) - g^{0}(\vec{\boldsymbol{R}}_{i}; \epsilon))^{2} d\epsilon}}{\sqrt{\frac{1}{N}\sum^{N}_{i}\int(g^{0}(\vec{\boldsymbol{R}}_{i}; \epsilon) - \overline{g^{0}}(\epsilon))^{2} d\epsilon}}\times100~,\label{eq:rmse}
\end{equation}
where $N$ is the number of structures in the target set, $\widetilde{g^{0}}(\vec{\boldsymbol{R}}_{i}; \epsilon)$ is the predicted and $g^{0}(\vec{\boldsymbol{R}}_{i}; \epsilon)$ is the target $0$ K eDOS of the $i$-th structure ($\vec{\boldsymbol{R}}_{i}$) and $\overline{g^{0}}(\epsilon)$ is the mean of the entire target set, given by
\begin{equation}
    \overline{g^{0}}(\epsilon) = \frac{1}{N}\sum^{N}_{i}g^{0}(\vec{\boldsymbol{R}}_{i}; \epsilon)~.\label{eq:edos_mean}
\end{equation}
As we can see, $\%$RMSE is the root mean square error (RMSE) of the prediction set normalized by the standard deviation (STD) of the target set.

In Fig. \ref{fig:edos}, we show the results of an ensemble of 10 best-performing models in the Bayesian hyperparameter optimization process. Even though we trained the whole eDOS, for convenience, we only show the more challenging region around the Fermi energy. The performance of the "ML used" model from Fig. \ref{fig:edos} on the test set is given in Table \ref{table:edos}. 
\begin{table}[h]
\caption{The performance of the "ML used" model from Fig. \ref{fig:edos} on the test set.}
\label{table:edos}
\centering
\begin{tabular}{|c|c|c|c|c|}
\hline
\text{model}&\text{MAE} & \text{RMSE} & \text{R}$^{2}$ & \text{\%RMSE} \\
\hline\hline
\text{ML used}&$0.93$ \text{$1/$eV}&$4.48$ \text{$1/$eV}&$0.99$ & $14.39\%$ \\
\hline
\end{tabular}
\end{table}
In \cite{PhysRevB.102.235130} it was shown that \%RMSE increases as $g_{b}$ decreases, and in that work $\%\text{RMSE}\approx10\%$ was achieved for $g_{b}=0.3$ eV and $\%\text{RMSE}\approx20\%$ for $g_{b}=0.1$ eV, while we achieve $\%\text{RMSE}=14.39\%$ for $g_{b}=0.01$ eV.
In Fig. \ref{fig:edos}, one can also see that the models sometimes predict negative values for the eDOS. Physically, this is impossible, but this is a consequence of the fact that the KRR model, by its construction, does not constrain eDOS predictions to be nonnegative. 

\subsection{Electronic free energy interatomic potential for finite $T_{el}$}
\label{subsec:finite_Tel}
\begin{figure}[h]
\centering
\includegraphics[width=\columnwidth]{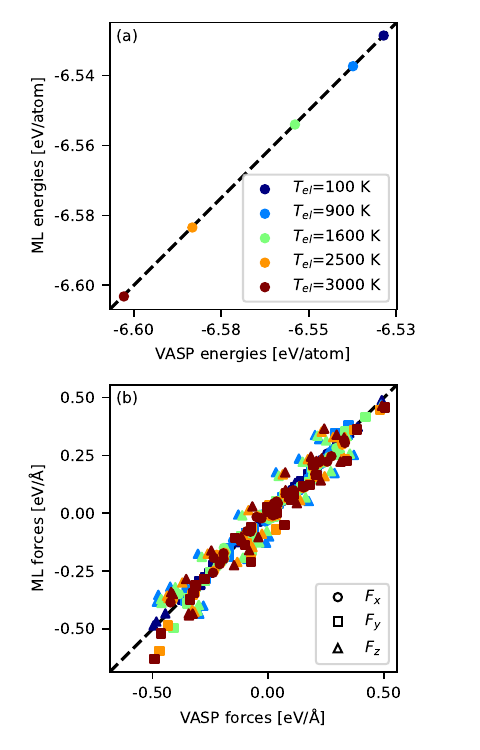}
\caption{(a) Correlation plot between VASP and full free energy ML potential energies and (b) correlation plot between VASP and free energy forces per component for five different electronic temperatures calculated for one $3\times3\times1$ structure from the test set.}
\label{fig:temperatures_subplots}
\end{figure}

By combining the models from \ref{subsec:E0_0K} and \ref{subsec:eDOS_0K} we construct the full ML model for electronic free energy \eqref{eq:energy} and corresponding forces \eqref{eq:force}.
To evaluate the performance of the full ML free energy potential we have randomly taken one $3\times3\times1$ structure from the test set and run VASP calculations for five different electronic temperatures. The correlation plots between the VASP and the full ML potential are shown in Fig. \ref{fig:temperatures_subplots}.
Performance metrics are reported in Table \ref{table:temperatures}. 
\begin{table}[H]
\caption{The performance of full ML potential evaluated on a one $3\times3\times1$ structure from the test set for five different electronic temperatures.}
\label{table:temperatures}
\centering
\begin{tabular}{|c|c|c|c|}
\hline
\text{quantity} & \text{MAE} & \text{RMSE} & \text{R}$^{2}$ \\
\hline\hline
\text{energy}&$0.13$ \text{meV/atom}&$0.17$ \text{meV/atom}&$1.00$\\
\hline
\text{F$_{x}$}&$12.80$ \text{meV/Å}&$17.29$ \text{meV/Å}&$0.99$\\
\hline
\text{F$_{y}$}&$27.24$ \text{meV/Å}&$39.84$ \text{meV/Å}&$0.97$\\
\hline
\text{F$_{z}$}&$55.04$ \text{meV/Å}&$71.29$ \text{meV/Å}&$0.93$\\
\hline
\end{tabular}
\end{table}
One can observe that the free electronic energy is well reproduced by our model for electronic temperatures in the range of $100-3000$ K, with errors still considerably lower than 1 meV/atom. Forces show slightly larger errors, up to an order of magnitude larger than errors for forces of $T_{el}=0$ ML model. Still, the errors are well below 100 meV/\AA~which is considered good for the kernel based methods \cite{PhysRevB.106.L121116}.
It can also be noticed in Fig. \ref{fig:temperatures_subplots} that as the electronic temperature increases, the forces become less accurate, which is also expected from the approximation in \eqref{eq:force}. It is also expected from the fact that as $T_{el}$ increases, DFT forces also have larger fluctuations. The possible solutions to this are given in \cite{PhysRevB.106.L121116} and the issue is addressed in more detail in Sec. \ref{sec:results_and_discussions}.

\subsection{Modelling ionic temperature}
\label{subsec:ionic_temperature}
In the previous subsections, we have presented the construction of the full ML model for the free electronic energy depending on electronic temperature ($T_{el}$) based on \eqref{eq:energy} and \eqref{eq:force}. To map a phase diagram, it is crucial to also include the effects of ionic temperature. While effects of ionic temperature can be modelled in several ways, we choose to use the stochastic self-consistent harmonic approximation (SSCHA) \cite{PhysRevB.96.014111, Monacelli_2021} that is often used to model CDW materials \cite{PhysRevLett.125.106101, doi:10.1021/acs.nanolett.0c00597} as it conveniently accounts for anharmonicity at a non-perturbative level including quantum and thermal fluctuations.
Previously, NbSe$_{2}$ was studied on the DFT level with SSCHA showing that significant anharmonic effects are present \cite{PhysRevLett.125.106101}, and these results can serve as benchmark for our ML models. Since SSCHA relies only on the energy and forces from any model, we can use it either with the $T_{el}=0$ K MACE model to reproduce previous DFT based results, or with the full free electronic energy ML model.

Most of the SSCHA parameters were kept at their default values. Initial harmonic dynamical matrices for the SSCHA calculations were obtained with Atomic Simulation Environment \cite{ISI:000175131400009, ase-paper}, with a finite difference step of 0.01 \AA.
Relaxation in SSCHA was performed with 500 configurations and, to generate ensembles, we have used 2000 configurations. To speed up calculations under laser conditions, the number of configurations was decreased to 250 and 1000, respectively. The meaningful factor was set to $10^{-5}$. Both the dynamical matrix and the structure minimization steps were set to 0.05.

All SSCHA calculations were performed with the random seed set to $0$, to ensure reproducibility.

\section{Results and discussions}
\label{sec:results_and_discussions}

In the following, we discuss the phonons dispersions as a function of $T_{el}$ and $T_{ion}$ and the CDW phase diagram as a function of both temperatures. The signature of the CDW phase is the existence of the imaginary phonon mode in the acoustic phonon branch between $\Gamma$ and M points of the Brillouin zone.
As discussed earlier, and as in previous works \cite{PhysRevLett.125.106101}, all calculations have been performed for $9\times9\times1$ unit cell that provides converged acoustic phonons in the $\Gamma$-M path. 

\subsection{Phonon dispersions as a function of $T_{el}$}
\label{subsec:phase_diagram_Tel}
\begin{figure}[ht]
\centering
\includegraphics[width=\columnwidth]{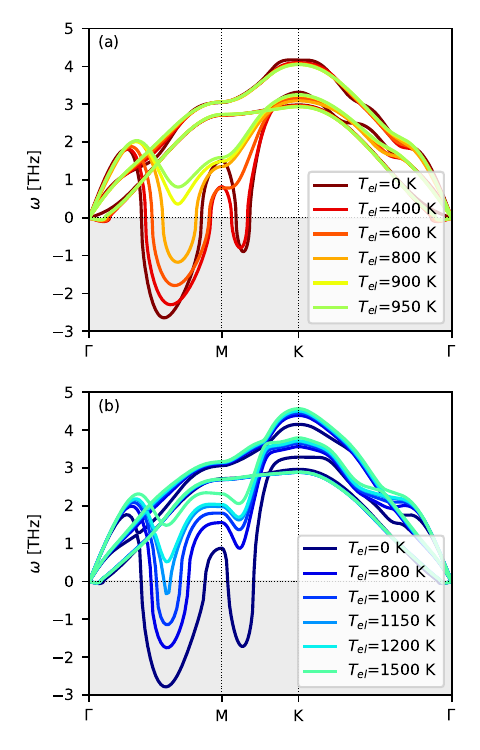}
\caption{(a) Acoustic phonon dispersions calculated in the harmonic approximation with DFT in $9\times9\times1$ supercell. (b) Acoustic phonon dispersions calculated in the harmonic approximation with the full free electronic ML potential in $9\times9\times1$ supercell.}
\label{fig:subplots_harmonic_acoustic}
\end{figure}

We have calculated the phonon dispersions in the harmonic approximation as a function of electronic temperature using the full free electronic enegy ML potential and VASP-DFT. 
The full phonon dispersions are given in Fig. S7, and the relevant acoustic phonons are given in Fig. \ref{fig:subplots_harmonic_acoustic}.
As discussed above, at $T_{el}=0$ K the two phonon dispersions (from DFT and our model) are comparable, as our MACE model is accurate. As expected, both DFT and our model predict the stabilization of the imaginary mode with increasing electronic temperature. Although the behaviour with increasing $T_{el}$ is similar, there are some noticable differences. First, one can notice that DFT predicts stabilization of the normal phase already for $T_{el}=900$ K, while in our model normal phase is obtained only for $T_{el}=1200$ K. There are two main reasons for this difference in the temperature of transition: starting approximations in the theoretical model \eqref{eq:energy} and \eqref{eq:force}, and errors associated with ML models. To check the accuracy of the approximation \eqref{eq:energy} in the theoretical model, we have plotted in Fig. S1 the free electronic energy along the CDW mode using the exact DFT eDOS. One can see that for $T_{el}=500$ K the approximation works well, but as $T_{el}$ increases there is a deviation and DFT gives higher phonon frequencies. Still, the temperature of the CDW transition between DFT and the approximation with the correct eDOS is similar. 

The second difference in phonon dispersions between the DFT and ML model is the location of the minimum of the imaginary mode (Kohn anomaly). In DFT, this minimum slightly shifts toward the M point as $T_{el}$ increases, while in the results of ML model it does not shift. However, this is an expected behaviour of the approximation in the theoretical model. Namely, in DFT the shift in position is due to the small changes in the parts of the Fermi surface that give rise to the Kohn anomaly. In the model, on the other hand, eDOS at $T_{el}=0$ K is used, and by construction it cannot capture the effects of changes in the electronic structure.
\begin{figure}[ht]
\center
\includegraphics[width=\columnwidth]{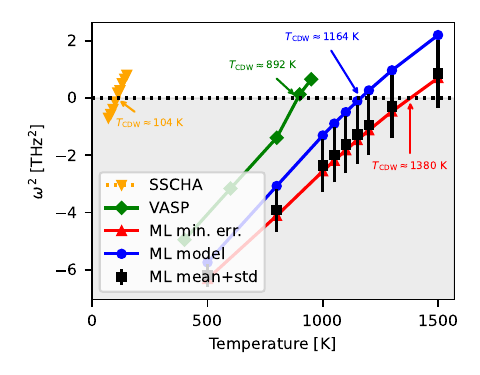}
\caption{Plot of the $\omega^{2}$ corresponding to the CDW instability along the $\Gamma$-M path versus temperature. It includes the VASP results, along with the results for the ensemble of 10 models explained earlier (in black). Also, in red we show the model with the minimal error \eqref{eq:rmse} in the Bayesian hyperparameter optimization process and in blue we show the model chosen out of this ensemble that was used for the calculations in this section. In orange we show the results of the SSCHA calculations with $T_{el}=0$ K.}
\label{fig:omega_squared}
\end{figure}

Since, near the CDW transition, the square of the phonon frequency $\omega^2$ can be approximated to depend linearly on the temperature \cite{PhysRevLett.125.106101}, in Fig. \ref{fig:omega_squared} we plot it for different models. We are plotting the $\omega^{2}$ of the instability along the $\Gamma$-M path. DFT predicts $T_{\rm CDW} \approx 892$ K and the ML model used for the calculations (blue in Fig. \ref{fig:omega_squared}) predicts $T_{\rm CDW}\approx1164$ K. This model is chosen such that out of the 10 best models according to the \%RMSE of the eDOS prediction on the validation set, it gives $T_{\rm CDW}$ closest to the one from DFT. The model with the lowest \%RMSE on the validation set gives $T_{\rm CDW}\approx1380$ K. Since we have constructed the dataset using the molecular dynamics to be used generally and not only for phonon calculations, no emphasis was given to particularly sample the configurations corresponding to the CDW mode. Optimizing the ML model solely based on the \%RMSE across all structures does not necessarily lead to optimal phonon predictions. While fine-tuning the dataset and model hyperparameters could improve the accuracy of phonons and $T_{\rm CDW}$ to match the limits of the underlying theoretical approximations, in this work we prioritized model generality.

Even with the said discrepancies, our model gives qualitatively correct results at orders of magnitude faster evaluation (for example, the calculations presented in Fig. \ref{fig:subplots_harmonic_acoustic}(a) took around a week using VASP with 64 MPI processes on a CPU cluster, while the calculations with our ML potential presented in Fig. \ref{fig:subplots_harmonic_acoustic}(b) took only a few minutes on a single CPU). It should also be noted that $T_{\rm CDW}$ and the corresponding CDW formation energy are sensitive to the calculation details and the DFT settings\,\cite{PhysRevB.106.245108,zheng22,hellgren2017}. In particular, the use of hybrid HSE exchange correlation functional can increase the CDW formation energy in NbSe$_2$ by an order of magnitude\,\cite{zheng22}, while $T_{\rm CDW}$ in TiSe$_2$ is enhanced from around 1100\,K to 1900\,K when PBE is replaced with HSE functional\,\cite{hellgren2017}. This shows that our model has smaller differences in $T_{\rm CDW}$ compared to DFT, than the DFT approximations itself.

\subsection{Phonon dispersions as a function of $T_{ion}$}
\label{subsec:phase_diagram_Tion}
In this subsection we report the acoustic phonon dispersions as a function of the ionic temperature using the SSCHA and keeping $T_{el}=0$ K, namely for these calculations, only the MACE potential for $E^{0}(\vec{\boldsymbol{R}}; 0)$ was used.
\begin{figure}[t]
\center
\includegraphics[width=\columnwidth]{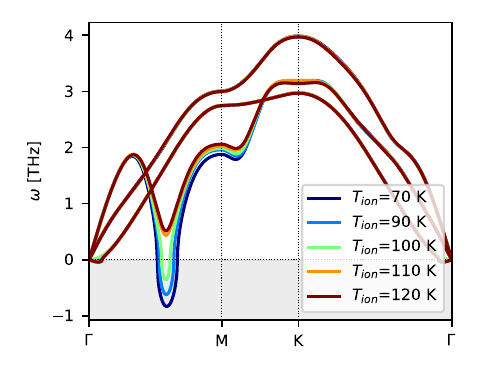}
\caption{Acoustic phonon dispersions calculated in the anharmonic approximation using the SSCHA with $T_{el}=0$ K and for different $T_{ion}$ in $9\times9\times1$ supercell.}
\label{fig:sscha_diff_temp}
\end{figure}
Anharmonic acoustic phonon dispersions are shown in Fig. \ref{fig:sscha_diff_temp} and are comparable with the DFT-SSCHA calculations in Ref.~\cite{PhysRevLett.125.106101}. 
Also, in Fig. \ref{fig:omega_squared} (orange line) it can be seen that our calculations give $T_{\rm CDW} \approx 104$ K, which is somewhat higher than the previous SSCHA calculations that predicted $T_{\rm CDW} \approx 70-80$\,K. Path-integral and \emph{ab initio} molecular dynamics simulations obtained $T_{\rm CDW}=75-100$\,K~\cite{zheng22}, which is consistent with our results. The spread in values for DFT based results stems from the difficulty in converging the results due to the computational complexity of DFT for such calculations, which is avoided in our approach. 
Our efficient ML model for modelling the CDW phase diagram therefore suggests that CDW in monolayer NbSe$_2$ is enhanced compared to the bulk case. In other words, our simulations are supporting the conclusions of experiments in Refs. \cite{Xi2015cdw,Lin2020cdw,azoury2023} which suggest $T_{\rm CDW}>100$ K, compared to the experiment from Ref. \cite{Ugeda2015cdw} which suggested $T_{\rm CDW}\approx 33$ K, close to the bulk value.

\subsection{CDW under nonequilibrium conditions}
\label{subsec:laser}

\begin{figure*}[ht]
\center
\includegraphics[width=2\columnwidth]{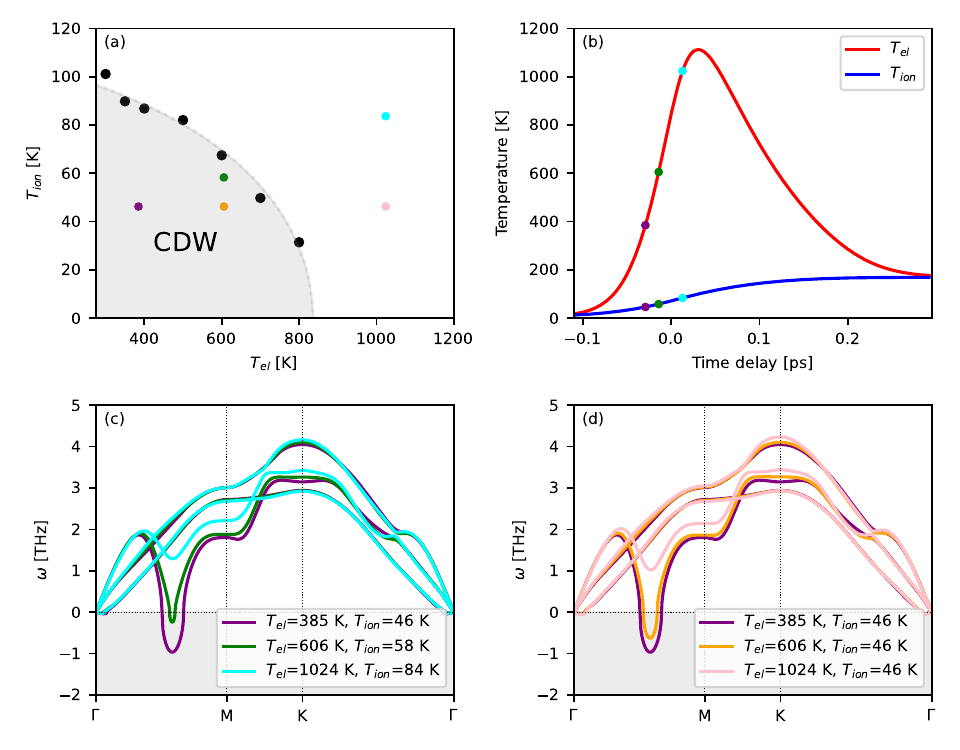}
\caption{(a) CDW order as a function of $T_{ion}$ and $T_{el}$, where the black circles indicate the pairs of $T_{ion}$ and $T_{el}$ after which the CDW instability disappears; the shaded area is bounded by fitting the black data points to the equation $T_{\rm CDW}(T_{el})=A\tanh{\sqrt{1-T_{el}/B}}$ \cite{PhysRevB.106.245108}, where $A=143$ K and $B=836$ K. (b) Results of a TTM for $T_{el}$ and $T_{ion}$; three pairs of $T_{el}$ and $T_{ion}$ are indicated, that are used in (c). (c) Acoustic phonon dispersions calculated using full interatomic ML potential with SSCHA for three pairs of $T_{el}$ and $T_{ion}$ indicated in (a) and (b). (d) Acoustic phonon dispersions calculated using the full interatomic ML potential with SSCHA for the same $T_{el}$ as in (c), but with $T_{ion}=46$ K kept fixed, as indicated in (a).}
\label{fig:laser}
\end{figure*}

An application of the full ML interatomic potential as obtained with Eqs.\,\eqref{eq:energy} and \eqref{eq:force} enables us to study the phase diagram as a function of both $T_{ion}$ and $T_{el}$ as shown in Fig.\,\ref{fig:laser}.
These calculations have been performed using the SSCHA with fixed $T_{el}$, for which the CDW phase still exists in the harmonic limit. For each $T_{el}$, we vary $T_{ion}$ until we get $T_{\rm CDW}$ in the same way as was done in Fig. \ref{fig:omega_squared}. In Fig.~\ref{fig:laser}(a) we report the ($T_{ion}$, $T_{el}$) phase diagram, which illustrates how the effects of electron and lattice entropy can melt the CDW phase, where obviously the latter is more dominant \cite{PhysRevLett.125.106101}. Nevertheless, two important conclusions regarding the role of electronic entropy emerge from this: (i) When trying to reach the quantitative values of thermal $T_{\rm CDW}$ it is important to choose appropriately small $T_{el}$ in the DFT calculations. (ii) Hot electrons (high $T_{el}$) can efficiently melt the CDW phase that originates from electron-phonon and phonon-phonon interactions. Namely, it is commonly believed that the ultrafast creation of hot electrons melts only the electron-induced order such as the excitonic insulator and the standard Peierls transition\,\cite{hellmann2012time,otto21}. For instance, sub-picosecond decay of the CDW signals in time-resolved measurements, which follows the increase of $T_{el}$, is usually considered a fingerprint of the purely electronic origin of the CDW\,\cite{hellmann2012time,otto21,monney16,payne2020}. Here we show that this can also be a fingerprint for phonon-related mechanism. Namely, soft Kohn anomaly in CDW-bearing TMDs is mostly induced by anisotropic electron-phonon scatterings, which means that anharmonicity is electron induced\,\cite{schobert24,varma1983,flicker2015charge,yoshida80,yoshiyama86}. In NbSe$_2$, instability of the acoustic phonon around the $2/3\, \Gamma \mathrm{M}$ point is driven by the electron scatterings between Nb-$d$ states around Fermi level\,\cite{flicker2015charge}, which are then diminished with elevated broadening or high $T_{el}$\,\cite{schobert24}.

To further corroborate this result, we perform TTM simulations using the DFT input parameters\,\cite{novko20,caruso22} (see supplementary note S4) and combine them with our ML model and SSCHA calculations [see Figs. \ref{fig:laser}(b)-(d)]. 
Considering that NbSe$_2$ is a metal in normal state, we assume that its electron thermalization time due to electron-electron scattering is very fast ($\sim 10$\,fs) and that the hot Fermi-Dirac distribution with $T_{el}$ is formed in a very early stage of dynamics\,\cite{caruso22}.
With this combination of techniques, we mimic the laser-induced time-resolved dynamics, as in time-resolved photoemission\,\cite{watanabe2022} and transient pump-probe optical spectroscopy\,\cite{payne2020,venturini2023}. In Fig. \ref{fig:laser}(b) we show the results of the TTM simulations (see also Fig. S9), where we can see a very fast dynamics of electron component of about 0.2\,ps, which agrees well with time-resolved photoemission results\,\cite{watanabe2022}, and it comes from the strong electron-phonon coupling of $\lambda = 1.3$. For laser fluence, we chose $F=0.1$\,J/m$^2$, which was found to be a critical value above which the CDW is melted in NbSe$_2$\,\cite{venturini2023}. Further, in Fig.\,\ref{fig:laser}(c) we show phonon dispersions of acoustic phonons with anharmonic corrections that correspond to three different snapshots, namely combinations of ($T_{el}$, $T_{ion}$), marked in panel (b). From this it is obvious that the CDW is melted by the time when $T_{el}$ reaches a peak, meaning slightly away from the zero time delay. In panel (d) we additionally show hardening of the CDW-related mode when $T_{ion}$ is kept fixed at the temperature of $46$\,K, while $T_{el}$ is raised according to TTM as in panels (b) and (c). By comparing these results and the one when $T_{ion}$ is varied in time according to TTM, it is obvious that the CDW is initially melted by hot electrons and that the timescale of CDW decay will follow sub-picosecond electron excitation timescale, i.e., $T_{el}$. Later on, in dynamics, the CDW is further quenched by the elevated $T_{ion}$ beyond $T_{\rm CDW} \approx 100$\,K, which supports the observation that the melting dynamics in NbSe$_2$ is long-lived\,\cite{venturini2023}.

\section{Conclusions}
\label{sec:conclusions}
In this work, we have constructed the first ML interatomic potential for the monolayer NbSe$_2$ with fairly low errors on energies and forces. The relatively long range of our potential made it possible to accurately capture acoustic phonons that are connected with CDW order. To model the effect of electronic temperature we have also constructed an ML model of electronic density of states that together with ML interatomic potential allows for accurate model of electronic free energies and forces, thus providing a full model capable of simulating NbSe$_2$ dynamics and phase diagram.

We have shown that our ML models can successfully predict the CDW phase diagram of NbSe$_2$ in agreement with DFT at orders of magnitude lower computational cost. With this model we obtain that the CDW transition temperature is 104\,K, which supports recent experimental claims that the CDW is enhanced for the monolayer NbSe$_2$.
We then used our model to investigate the CDW phase diagram as a function of both electronic and ion temperature that arise in the non-thermal conditions such as under the short laser pulse. We show that elevated electronic temperatures efficiently melt the electron-phonon-driven CDW order, which is then long-lived as the ionic temperature equilibrates due to the electron-phonon coupling. Our results show that melting of CDW in time-resolved measurements cannot be exclusively assigned to the purely electronic origin of the CDW as often assumed. This can be especially useful for CDW-bearing systems, where time-resolved measurements are often utilized for the disentanglement of purely electronic and phonons origins to the formation of the CDW, such as for understanding the timescales of excitonic insulator and Jahn-Teller mechanisms in TiSe$_2$\,\cite{porer14} and similar TMDs.

\section*{Data availability}
All DFT training data and ML models, as well as calculation scripts, are available in Zenodo with the identifier 10.5281/zenodo.15125087.

\section*{Code availability}
All code versions and modifications in the codes are available in Zenodo with the identifier 10.5281/zenodo.15125087.

\section*{Acknowledgments}
\label{sec:acknowledgments}
LB would like to express his gratitude to Juraj Ovčar for many useful discussions. This work has been supported by Croatian Science Foundation under the project UIP-2020-02-5675, the European Regional Development Fund within the ``Center of Excellence for Advanced Materials and Sensing Devices'' (Grant No. KK.01.1.1.01.0001), and COST Action CA22154 - Data-driven Applications towards the Engineering of functional Materials: an Open Network (DAEMON) supported by COST (European Cooperation in Science and Technology). FG acknowledges financial support from the EMPEROR Project, CUP E93C24001040001, sponsored by the National Quantum Science and Technology Institute (Spoke 5), Grant No. PE00000023, funded by the European Union -- NextGeneration EU. CBM acknowledges funding from the Swiss National
Science Foundation (SNSF) under grant number 217837. DN acknowledges financial support from the project ``Podizanje znanstvene izvrsnosti Centra za napredne laserske tehnike (CALTboost)'' financed by the European Union through the National Recovery and Resilience Plan 2021-2026 (NRPP). 

\section*{Author Contributions}
LB has trained all ML models and performed all calculations with them under the supervision of IL. IL and LB performed DFT calculations for the training dataset. DN performed TTM calculations. IL designed the study with input from FG and DN. FG and CBM helped in the construction of electronic free energy model.
LB, DN and IL wrote the manuscript draft. All authors provided comments on the manuscript. All authors read and approved the final manuscript.

\section*{Competing interests}
The authors declare no competing interests.

\bibliography{ref}

\end{document}


\title{Supplemental Material:\\ Machine learning model for efficient nonthermal tuning of the charge density wave in monolayer NbSe$_2$}

\author{
    Luka Benić$^{1}$, Federico Grasselli$^{2,3}$, Chiheb Ben Mahmoud$^{4}$, Dino Novko$^{5, *}$ and Ivor Lončarić$^{1,\dagger}$ \\
    $^{1}$\textit{Ruđer Bošković Institute, Zagreb, Croatia} \\
    $^{2}$\textit{Dipartimento di Scienze Fisiche, Informatiche e Matematiche, Universit\`a degli Studi di Modena e Reggio Emilia, Modena, Italy} \\
    $^{3}$\textit{CNR NANO S3, Modena, Italy} \\
    $^{4}$\textit{Inorganic Chemistry Laboratory, Department of Chemistry, University of Oxford, Oxford OX1 3QR, United Kingdom} \\
    $^{5}$\textit{Centre for Advanced Laser Techniques, Institute of Physics, Zagreb, Croatia} \\
    $^{*}$\href{mailto:dino.novko@gmail.com} {dino.novko@gmail.com}, $^{\dagger}$\href{mailto:ivor.loncaric@gmail.com}{ivor.loncaric@gmail.com} 
}

\maketitle
\thispagestyle{empty}

\newpage
\section{Approximation for the Helmholtz electronic free energy and the ionic forces}
\label{S_approximation}

\setcounter{page}{1} 
\pagenumbering{arabic}

We report the derivation of the approximation used for simulating the finite electronic temperature effects. The idea is to use the $0$ K density functional theory data to construct an approximation for the Helmholtz electronic free energy to be able to efficiently simulate finite electronic temperature effects. It is a modified derivation of the approximation developed in Ref.  \cite{PhysRevB.106.L121116}. Firstly, we define the notation we will use. We start by considering an atomic system with $N_{ion}$ ions and $N_{el}$ electrons, where $\vec{\boldsymbol{R}}=\{\boldsymbol{R}_{1},...,\boldsymbol{R}_{N_{ion}}\}$ represents the set of ionic coordinates and we let $T_{el}$ to be the electronic temperature of the system. Fermi-Dirac distribution is given by
\begin{equation}
    f(x)=\frac{1}{e^{x}+1}~.\label{eq:S1}
\end{equation}

In the following derivations, the temperatures in the superscripts indicate to which temperatures do the eigenvalues that are used to construct desired quantities correspond, and the temperatures in parentheses indicate at which temperatures the desired quantities are evaluated.

We define occupation numbers as
\begin{equation}
    f^{T_{1}}_{i}(T_{2}) \stackrel{\eqref{eq:S1}}{=} f\left(\frac{\epsilon^{T_{1}}_{i}(\vec{\boldsymbol{R}}) - \mu^{T_{1}}(\vec{\boldsymbol{R}}; T_{2})}{k_{B}T_{2}}\right)~,\label{eq:S2}
\end{equation}
where $k_{B}$ is the Boltzmann constant. In \eqref{eq:S2} \textit{i}-th occupation number $f^{T_{1}}_{i}(T_{2})$ evaluated at electronic temperature $T_{2}$ is determined by the \textit{i}-th Kohn-Sham eigenenergy $\epsilon^{T_{1}}_{i}(\vec{\boldsymbol{R}})$ corresponding to $T_{1}$ and the chemical potential (Fermi level) $\mu^{T_{1}}(\vec{\boldsymbol{R}}; T_{2})$ evaluated at $T_{2}$ and constructed out of the $T_{1}$ eigenenergies. The chemical potential is determined by the constraint of conservation of the number of electrons in the system
\begin{equation}
    N_{el} = \sum_{i}f^{T_{1}}_{i}(T_{2})~\forall\hspace{1mm}T_{1}, T_{2}~,\label{eq:S3}
\end{equation}
where the summation over \textit{i} represents the summation over the spin projections \textit{s}, bands \textit{n} and k-points $\boldsymbol{k}$. Furthermore, in the limit $T_{2}\rightarrow0^{+}$ we have
\begin{equation}
    \mu^{T_{1}}(\vec{\boldsymbol{R}}; 0) = \epsilon^{T_{1}}_{F}(\vec{\boldsymbol{R}})~,\label{eq:S4}
\end{equation}
where we call $\epsilon^{T_{1}}_{F}(\vec{\boldsymbol{R}})$ the Fermi energy of the system at the electronic temperature $T_{1}$, which in the limit $T_{1}\rightarrow0^{+}$ becomes the real Fermi energy of the system $\epsilon^{0}_{F}(\vec{\boldsymbol{R}})$.

Next, we define the electronic density of states (eDOS) at a electronic temperature \textit{T} as
\begin{equation}
    g^{T}(\vec{\boldsymbol{R}}; \epsilon) = \sum_{i}\delta(\epsilon-\epsilon^{T}_{i}(\vec{\boldsymbol{R}}))~,\label{eq:S5}
\end{equation}
where $\delta(x)$ is the Dirac delta distribution. Using \eqref{eq:S5} we have the standard correspondence
\begin{equation}
    \sum_{i}\rightarrow\int g^{T}(\vec{\boldsymbol{R}}; \epsilon)d\epsilon~.\label{eq:S6}
\end{equation}
Using \eqref{eq:S6} the equation \eqref{eq:S3} can be written as
\begin{equation}
    N_{el} = \int g^{T_{1}}(\vec{\boldsymbol{R}}; \epsilon)f\left(\frac{\epsilon - \mu^{T_{1}}(\vec{\boldsymbol{R}}; T_{2})}{k_{B}T_{2}}\right)d\epsilon~.\label{eq:S7}
\end{equation}
The fermionic entropy in this notation is defined as
\begin{align}
    S^{T_{1}}(\vec{\boldsymbol{R}}; T_{2}) = -k_{B}\sum_{i}\left[f^{T_{1}}_{i}(T_{2})\ln(f^{T_{1}}_{i}(T_{2})) + (1-f^{T_{1}}_{i}(T_{2}))\ln(1-f^{T_{1}}_{i}(T_{2}))\right]~.\label{eq:S8}
\end{align}
For the electronic density, we have
\begin{equation}
    \rho^{T_{1}}(\boldsymbol{r};T_{2}) = \sum_{i} \rho^{T_{1}}_{i}(\boldsymbol{r};T_{2}) = \sum_{i}f^{T_{1}}_{i}(T_{2})|\phi^{T_{1}}_{i}(\boldsymbol{r})|^{2}~,\label{eq:S9}
\end{equation}
where $\phi^{T_{1}}_{i}(\boldsymbol{r})$ is the \textit{i}-th Kohn-Sham orbital at the electronic temperature $T_{1}$ and $\boldsymbol{r}=r_{x}\boldsymbol{\hat{x}}+r_{y}\boldsymbol{\hat{y}}+r_{z}\boldsymbol{\hat{z}}$.

The Helmholtz electronic free energy of a system at the electronic temperature $T_{el}$, with fixed number of electrons $N_{el}$ and fixed volume $V$ of the system is given by \cite{PhysRev.137.A1441}
\begin{equation}
    F(\vec{\boldsymbol{R}}; N_{el}, V, T_{el}) \equiv F(\vec{\boldsymbol{R}}; T_{el}) = E^{T_{el}}(\vec{\boldsymbol{R}}; T_{el}) - T_{el}S^{T_{el}}(\vec{\boldsymbol{R}}; T_{el})~,\label{eq:S10}
\end{equation}
where $E^{T_{el}}(\vec{\boldsymbol{R}}; T_{el})$ is the standard DFT energy defined as \cite{giustino2014materials}
\begin{equation}
    E^{T_{el}}(\vec{\boldsymbol{R}}; T_{el}) = E_{b}^{T_{el}}(\vec{\boldsymbol{R}}; T_{el}) - E^{T_{el}}_{dc}(\vec{\boldsymbol{R}}; T_{el}) + E_{nn}(\vec{\boldsymbol{R}})~.\label{eq:S11}
\end{equation}
In \eqref{eq:S11} the first, so-called \textit{band-energy}, term, is given by
\begin{equation}
    E^{T_{el}}_{b}(\vec{\boldsymbol{R}}; T_{el}) = \sum_{i}f^{T_{el}}_{i}(T_{el})\epsilon^{T_{el}}_{i}(\vec{\boldsymbol{R}})~.\label{eq:S12}
\end{equation}
The second term, the so-called \textit{double-counting} term, is given by
\begin{align}
    E^{T_{el}}_{dc}(\vec{\boldsymbol{R}}; T_{el}) = \frac{1}{2}\iint&\frac{\rho^{T_{el}}(\boldsymbol{r};T_{el})\rho^{T_{el}}(\boldsymbol{r}';T_{el})}{|\boldsymbol{r}-\boldsymbol{r}'|}d\boldsymbol{r}d\boldsymbol{r}' - E_{XC}[\rho^{T_{el}}(\boldsymbol{r};T_{el})](T_{el}) \nonumber \\
    & \hspace{-0.75cm} + \int V_{XC}[\rho^{T_{el}}(\boldsymbol{r};T_{el})](\boldsymbol{r}; T_{el})\rho^{T_{el}}(\boldsymbol{r};T_{el})d\boldsymbol{r}~,\label{eq:S13}
\end{align}
where $E_{XC}[\rho^{T_{el}}(\boldsymbol{r};T_{el})](T_{el})$ is the so-called \textit{exchange-correlation} (XC) functional and
\begin{equation}
    V_{XC}[\rho^{T_{el}}(\boldsymbol{r};T_{el})](\boldsymbol{r}; T_{el}) = \frac{\delta E_{XC}[\rho^{T_{el}}(\boldsymbol{r};T_{el})](T_{el})}{\delta\rho^{T_{el}}(\boldsymbol{r};T_{el})}~\label{eq:S14}
\end{equation}
is the XC potential. Finally, the last term in \eqref{eq:S11} is the electrostatic energy between the ions and is independent of $T_{el}$.

We start by taking the difference between the Helmholtz energies \eqref{eq:S10} at the two different temperatures, where some finite reference temperature $T_{1}$ is taken to make the derivations more consistent. All of the results will be based at $T_{1}\rightarrow0^{+}$ and at the end we will let $T_{1}\rightarrow0^{+}$ everywhere. So, we start with
\begin{align}
    & F(\vec{\boldsymbol{R}}; T_{2}) - F(\vec{\boldsymbol{R}}; T_{1}) = E^{T_{2}}_{b}(\vec{\boldsymbol{R}}; T_{2}) - E^{T_{1}}_{b}(\vec{\boldsymbol{R}}; T_{1}) - E^{T_{2}}_{dc}(\vec{\boldsymbol{R}}; T_{2}) + E^{T_{1}}_{dc}(\vec{\boldsymbol{R}}; T_{1}) \nonumber \\
    & \hspace{2cm} + \cancel{E_{nn}(\vec{\boldsymbol{R}})} - \cancel{E_{nn} (\vec{\boldsymbol{R}})} - T_{2}S^{T_{2}}(\vec{\boldsymbol{R}}; T_{2}) + T_{1}S^{T_{1}}(\vec{\boldsymbol{R}}; T_{1})~.\label{eq:S15}
\end{align}
Now, the approximation consists of using a similar idea as in the Ref. \cite{PhysRevB.18.7165}. However, keeping in mind that the results from \cite{PhysRevB.18.7165} are obtained for $T_{el}=0$ K. The idea there was that even at $T_{el}=0$ K one can introduce occupational numbers like in \eqref{eq:S9} as a mathematical tool that enables to calculate variations with respect to occupational numbers, and at the end of the calculations one lets those occupational numbers follow the Fermi-Dirac distribution at $T_{el}=0$ K, or in other words, occupational numbers can take only the values of $0$ and $1$. The main result of that paper is the so-called Janak's theorem, which states that from \eqref{eq:S11} at $T_{el}=0$ K one has
\begin{align}
    \left.\frac{\delta E^{T_{el}}(\vec{\boldsymbol{R}};T_{el})}{\delta f^{T_{el}}_{i}(T_{el})}\right|_{T_{el}=0}=\epsilon^{0}_{i}(\vec{\boldsymbol{R}})~.\label{eq:S16}
\end{align}
It is worth mentioning that Janak's theorem has its thermal version \cite{10.1143/PTP.111.339}, but this is not what this approximation aims to use. The variation with respect to occupational numbers can be traced to the following idea
\begin{align}
    \psi_{\vec{\boldsymbol{R}}}(\vec{\boldsymbol{r}};0)\iff\rho^{0}({\boldsymbol{r}};0)\stackrel{\eqref{eq:S9}}{\iff}\{f^{0}_{i}(0)\}, \{\phi^{0}_{i}(\boldsymbol{r})\}~,\label{eq:S17}
\end{align}
where $\psi_{\vec{\boldsymbol{R}}}(\vec{\boldsymbol{r}};0)$ is the exact electronic wave function at $0$ K. The first equivalence in \eqref{eq:S17} is due to the Hohenberg-Kohn theorem \cite{PhysRev.136.B864}. In \eqref{eq:S17} we want to emphasize that once we have density functional description of the quantum many-body problem at $0$ K given by the Hohenberg-Kohn theorem, we can formally describe it using the occupational numbers and Kohn-Sham orbitals, as implied by \eqref{eq:S9}, where we impose that those occupational numbers obey the Fermi-Dirac distribution at $0$ K. This is the reasoning behind taking the variations with respect to occupational numbers in \eqref{eq:S16}. It also allows one to treat the variations with respect to occupational numbers and Kohn-Sham orbitals independently. So, the idea is to treat the occupational numbers introduced in \cite{PhysRevB.18.7165} as objects with physical meaning to capture the effects of changing the electronic temperature and to construct an approximation for the Helmholtz electronic free energy. Using only the information we have at $T_{1}\rightarrow0^{+}$, we perform the following set of approximations
\begin{align}
&\hspace{0.5cm}\rho^{T_{2}}(\boldsymbol{r}; T_{2}) \approx\rho^{T_{1}}(\boldsymbol{r}; T_{1}) + \sum_{i}\frac{\delta\rho^{T_{1}}(\boldsymbol{r}; T)}{\delta f^{T_{1}}_{i}(T)}\Bigg|_{T=T_{1}}(f^{T_{1}}_{i}(T_{2})-f^{T_{1}}_{i}(T_{1}))~, \label{eq:S18}
\\
&E^{T_{2}}_{b}(\vec{\boldsymbol{R}};T_{2})\approx E^{T_{1}}_{b}(\vec{\boldsymbol{R}};T_{1})+\sum_{i}\frac{\delta E^{T_{1}}_{b}(\vec{\boldsymbol{R}}; T)}{\delta f^{T_{1}}_{i}(T)}\Bigg|_{T=T_{1}}(f^{T_{1}}_{i}(T_{2})-f^{T_{1}}_{i}(T_{1}))~, \label{eq:S19}
\\
&E^{T_{2}}_{dc}(\vec{\boldsymbol{R}};T_{2})\approx E^{T_{1}}_{dc}(\vec{\boldsymbol{R}};T_{1})+\sum_{i}\frac{\delta E^{T_{1}}_{dc}(\vec{\boldsymbol{R}}; T)}{\delta f^{T_{1}}_{i}(T)}\Bigg|_{T=T_{1}}(f^{T_{1}}_{i}(T_{2})-f^{T_{1}}_{i}(T_{1}))~, \label{eq:S20}
\\
&\hspace{4cm}T_{2}S^{T_{2}}(\vec{\boldsymbol{R}};T_{2})\approx T_{2}S^{T_{1}}(\vec{\boldsymbol{R}};T_{2})~. \label{eq:S21}
\end{align}
The approximations given by \eqref{eq:S18}-\eqref{eq:S21} tell us that we approximate the quantites in \eqref{eq:S18}-\eqref{eq:S20} using the perturbations in occupational numbers caused by change in temperature. 
\\
From \eqref{eq:S17} and \eqref{eq:S18} we have
\begin{align}
    & \hspace{0.5cm}\rho^{T_{2}}(\boldsymbol{r}; T_{2}) \approx \rho^{T_{1}}(\boldsymbol{r}; T_{1}) + \sum_{i}\frac{\delta\rho^{T_{1}}(\boldsymbol{r}; T)}{\delta f^{T_{1}}_{i}(T)}\Bigg|_{T=T_{1}}(f^{T_{1}}_{i}(T_{2})-f^{T_{1}}_{i}(T_{1})) \nonumber \\
    & \hspace{1.75cm} = \rho^{T_{1}}(\boldsymbol{r}; T_{1}) + \sum_{i}(f^{T_{1}}_{i}(T_{2})-f^{T_{1}}_{i}(T_{1}))|\phi^{T_{1}}_{i}(\boldsymbol{r})|^{2} \nonumber \\
    & \hspace{0.75cm} = \cancel{\rho^{T_{1}}(\boldsymbol{r}; T_{1})} + \underbrace{\sum_{i}f^{T_{1}}_{i}(T_{2})|\phi^{T_{1}}_{i}(\boldsymbol{r})|^2}_{\stackrel{\eqref{eq:S9}}{=\rho^{T_{1}}}(\boldsymbol{r};T_{2})} - \cancel{\underbrace{\sum_{i}f^{T_{1}}_{i}(T_{1})|\phi^{T_{1}}_{i}(\boldsymbol{r})|^2}_{\stackrel{\eqref{eq:S9}}{=}\rho^{T_{1}}(\boldsymbol{r};T_{1})}} \nonumber \\
    & \hspace{3.25cm} \implies \rho^{T_{2}}(\boldsymbol{r}; T_{2}) \approx \rho^{T_{1}}(\boldsymbol{r};T_{2})~.\label{eq:S22}
\end{align}
Now, using \eqref{eq:S18}-\eqref{eq:S21} in \eqref{eq:S15} we have
\begin{align}
    & \hspace{1.25cm} F(\vec{\boldsymbol{R}};T_{2})-F(\vec{\boldsymbol{R}};T_{1})\approx\sum_{i}\frac{\delta E^{T_{1}}_{b}(\vec{\boldsymbol{R}}; T)}{\delta f^{T_{1}}_{i}(T)}\Bigg|_{T=T_{1}}(f^{T_{1}}_{i}(T_{2})-f^{T_{1}}_{i}(T_{1})) \nonumber \\
    & -\sum_{i}\frac{\delta E^{T_{1}}_{dc}(\vec{\boldsymbol{R}}; T)}{\delta f^{T_{1}}_{i}(T)}\Bigg|_{T=T_{1}}(f^{T_{1}}_{i}(T_{2})-f^{T_{1}}_{i}(T_{1}))-T_{2}S^{T_{1}}(\vec{\boldsymbol{R}};T_{2})+T_{1}S^{T_{1}}(\vec{\boldsymbol{R}};T_{1})~.\label{eq:S23}
\end{align}
Also, we know that
\begin{align}
    \frac{\delta E_{nn}(\vec{\boldsymbol{R}})}{\delta f^{T_{1}}_{i}(T)}=0~.\label{eq:S24}
\end{align}
Using \eqref{eq:S24} in \eqref{eq:S23} we have
\begin{align}
    & F(\vec{\boldsymbol{R}};T_{2})-F(\vec{\boldsymbol{R}};T_{1})\approx\sum_{i}\frac{\delta \left(E^{T_{1}}_{b}(\vec{\boldsymbol{R}}; T)-E^{T_{1}}_{dc}(\vec{\boldsymbol{R}}; T)+E_{nn}(\vec{\boldsymbol{R}})\right)}{\delta f^{T_{1}}_{i}(T)}\Bigg|_{T=T_{1}}(f^{T_{1}}_{i}(T_{2})-f^{T_{1}}_{i}(T_{1})) \nonumber \\
    & \hspace{4.75cm}-T_{2}S^{T_{1}}(\vec{\boldsymbol{R}};T_{2})+T_{1}S^{T_{1}}(\vec{\boldsymbol{R}};T_{1}) \nonumber \\
    & \hspace{0.85cm}\stackrel{\eqref{eq:S11}}{=}\sum_{i}\frac{\delta E^{T_{1}}(\vec{\boldsymbol{R}};T)}{\delta f^{T_{1}}_{i}(T)}\Bigg|_{T=T_{1}}(f^{T_{1}}_{i}(T_{2})-f^{T_{1}}_{i}(T_{1}))-T_{2}S^{T_{1}}(\vec{\boldsymbol{R}};T_{2})+T_{1}S^{T_{1}}(\vec{\boldsymbol{R}};T_{1}) \nonumber \\
    & \hspace{1.85cm}\stackrel{\eqref{eq:S16}}{=}\sum_{i}\epsilon^{T_{1}}_{i}(\vec{\boldsymbol{R}})(f^{T_{1}}_{i}(T_{2})-f^{T_{1}}_{i}(T_{1}))-T_{2}S^{T_{1}}(\vec{\boldsymbol{R}};T_{2})+T_{1}S^{T_{1}}(\vec{\boldsymbol{R}};T_{1})~.\label{eq:S25}
\end{align}
In conclusion, using \eqref{eq:S10} and \eqref{eq:S25} we approximate the Helmholtz electronic free energy evaluated at a finite electronic temperature $T_{2}$ as
\begin{align}
    F(\vec{\boldsymbol{R}}; T_{2}) \approx E^{T_{1}}(\vec{\boldsymbol{R}}; T_{1})+\sum_{i}\epsilon^{T_{1}}_{i}(\vec{\boldsymbol{R}})(f^{T_{1}}_{i}(T_{2})-f^{T_{1}}_{i}(T_{1})) - T_{2}S^{T_{1}}(\vec{\boldsymbol{R}}; T_{2})~.\label{eq:S26}
\end{align}
Furthermore, as we have already mentioned, $T_{1}$ was used just to have a more mathematically consistent derivation, so we can finally write everything using $T_{1}\rightarrow0^{+}$, where we also let $T_{2}=T_{el}$
\begin{align}
    F(\vec{\boldsymbol{R}};T_{el})\approx E^{0}(\vec{\boldsymbol{R}};0)+\sum_{i}\epsilon^{0}_{i}(\vec{\boldsymbol{R}})(f^{0}_{i}(T_{el})-f^{0}_{i}(0)) - T_{el}S^{0}(\vec{\boldsymbol{R}}; T_{el})~.\label{eq:S27}
\end{align}
We can write \eqref{eq:S26} as
\begin{align}
    F(\vec{\boldsymbol{R}}; T_{2}) \approx E^{T_{1}}(\vec{\boldsymbol{R}}; T_{1}) + \Delta F(\vec{\boldsymbol{R}};T_{1},T_{2})~,\label{eq:S28}
\end{align}
where we have defined
\begin{align}
    \Delta F(\vec{\boldsymbol{R}};T_{1},T_{2}) = \sum_{i}\epsilon^{T_{1}}_{i}(\vec{\boldsymbol{R}})(f^{T_{1}}_{i}(T_{2})-f^{T_{1}}_{i}(T_{1}))- T_{2}S^{T_{1}}(\vec{\boldsymbol{R}}; T_{2})~.\label{eq:S29}
\end{align}
Or, using $T_{1}\rightarrow0^{+}$ and the definitions from \eqref{eq:S27} we have
\begin{align}
    F(\vec{\boldsymbol{R}};T_{el})\approx E^{0}(\vec{\boldsymbol{R}};0)+\Delta F(\vec{\boldsymbol{R}};0,T_{el})~.\label{eq:S30}
\end{align}
From \eqref{eq:S6}, \eqref{eq:S8}, and \eqref{eq:S26} we have
\begin{align}
    & F(\vec{\boldsymbol{R}}; T_{2}) \approx E^{T_{1}}(\vec{\boldsymbol{R}}; T_{1}) + \int\epsilon g^{T_{1}}(\vec{\boldsymbol{R}}; \epsilon)\left[ f\left(\frac{\epsilon - \mu^{T_{1}}(\vec{\boldsymbol{R}}; T_{2})}{k_{B}T_{2}}\right) - f\left(\frac{\epsilon - \mu^{T_{1}}(\vec{\boldsymbol{R}}; T_{1})}{k_{B}T_{1}}\right) \right] d\epsilon \nonumber \\
    & \hspace{4cm} + k_{B}T_{2}\int g^{T_{1}}(\vec{\boldsymbol{R}}; \epsilon)s\left(\frac{\epsilon - \mu^{T_{1}}(\vec{\boldsymbol{R}}; T_{2})}{k_{B}T_{2}}\right) d\epsilon~,\label{eq:S31}
\end{align}
where $s(x)$ is defined as
\begin{align}
    s(x) = f(x) \ln \left( f(x) \right) + (1-f(x))\ln(1-f(x))~.\label{eq:S32}
\end{align}
Or, using $T_{1}\rightarrow0^{+}$ and the definitions from \eqref{eq:S27}, along with \eqref{eq:S4}, we have
\begin{align}
    & F(\vec{\boldsymbol{R}};T_{el})\approx E^{0}(\vec{\boldsymbol{R}};0)+\int\epsilon g^{0}(\vec{\boldsymbol{R}}; \epsilon)\left[ f\left(\frac{\epsilon - \mu^{0}(\vec{\boldsymbol{R}}; T_{2})}{k_{B}T_{2}}\right) - \theta(\epsilon^{0}_{F}(\vec{\boldsymbol{R}})-\epsilon) \right] d\epsilon \nonumber \\
    & \hspace{3.25cm} + k_{B}T_{2}\int g^{0}(\vec{\boldsymbol{R}}; \epsilon)s\left(\frac{\epsilon - \mu^{0}(\vec{\boldsymbol{R}}; T_{2})}{k_{B}T_{2}}\right) d\epsilon~,\label{eq:S33}
\end{align}
where $f^{0}(0)=\theta(\epsilon^{0}_{F}(\vec{\boldsymbol{R}})-\epsilon)$ was used.

In essence, this approximation exploits the idea from \cite{PhysRevB.18.7165} where the occupation numbers are let to be fractional and have been used as a mathematical tool to derive Janak's theorem. But in this approximation they are treated as valid physical objects and are used to simulate changes in temperature and to construct an approximation for the Helmholtz electronic free energy. Inspecting the presented derivations, one can say that the approximation developed is based on the following approximation $f^{T_{2}}_{i}(T_{2})\approx f^{T_{1}}_{i}(T_{2})$, which, as already said, means that occupational numbers introduced in the Janak's theorem were used to simulate the effects of change in temperature. In conclusion, thermal effects are included by extrapolating the $0$ K density functional theory data to finite temperatures using the constructed approximation for the Helmholtz electronic free energy. The main difference between this derivation of \eqref{eq:S33} and the one from \cite{PhysRevB.106.L121116} is in the fact that here we did not use $f^{T_{2}}_{i}(T_{2})$ quantities explicitly, for which we do not have any information in this model.

\begin{figure}
    \centering
    \includegraphics[width=\columnwidth]{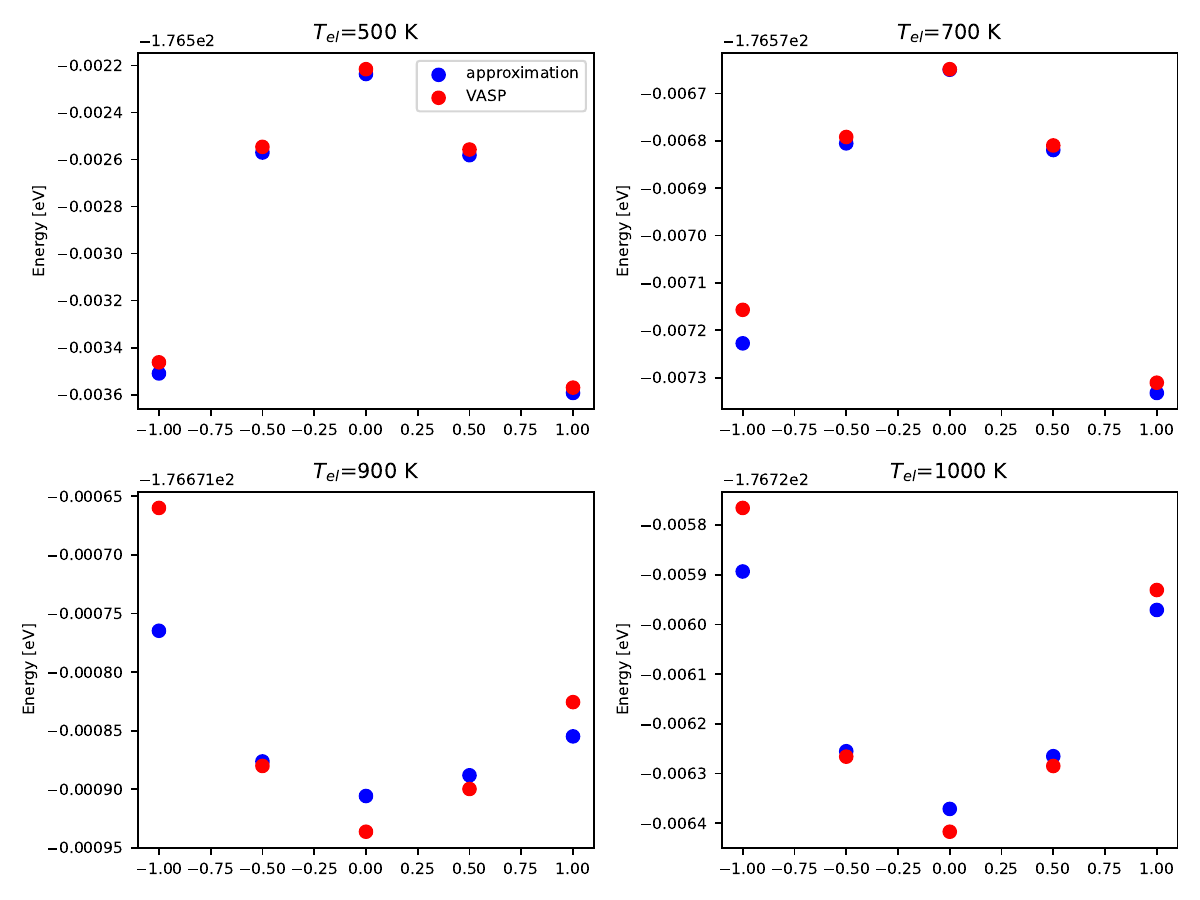}
    \caption{Comparison of the approximation \eqref{eq:S33} and the VASP results for the Helmholtz electronic free energy for four different electronic temperatures $T_{el}$. Calculations were performed along the $\boldsymbol{q}=2/3~\Gamma$M CDW phonon mode in $3\times3\times1$ supercell. For these calculations every eDOS was constructed using the same hyperparameters as in the main text. Namely, $g_{b}=0.01$ eV and $d\epsilon=0.00025$ eV.}
    \label{fig:subplots_approximation}
\end{figure}

\begin{figure}
    \centering
    \includegraphics[width=\columnwidth]{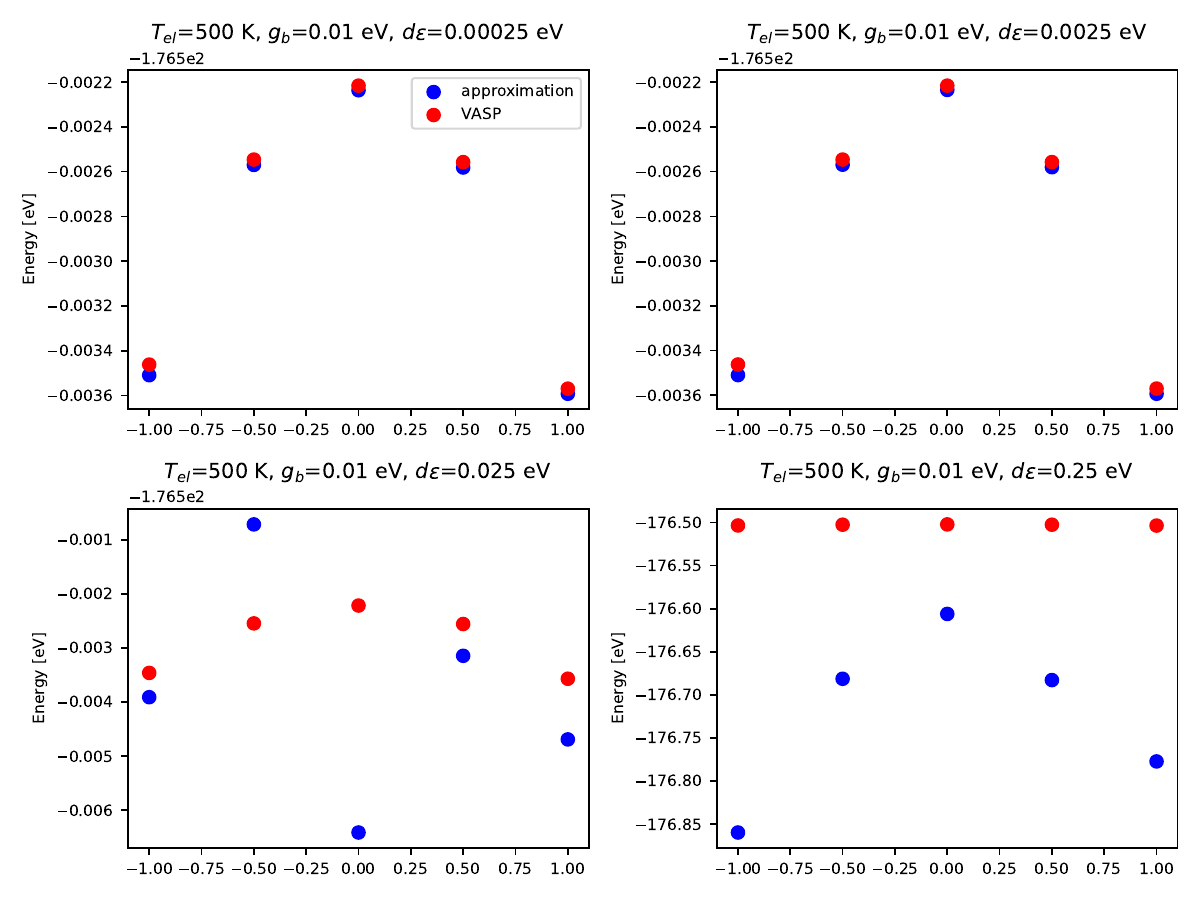}
    \caption{Comparison between the approximation \eqref{eq:S33} and the VASP results for the Helmholtz electronic free energy calculated for one $3\times3\times1$ structure along the $\boldsymbol{q}=2/3~\Gamma$M CDW phonon mode for $T_{el}=500$ K, $g_{b}=0.01$ eV and four different $d\epsilon$ hyperparameters.}
    \label{fig:diff_d_eps}
\end{figure}

\begin{figure}
    \centering
    \includegraphics[width=\columnwidth]{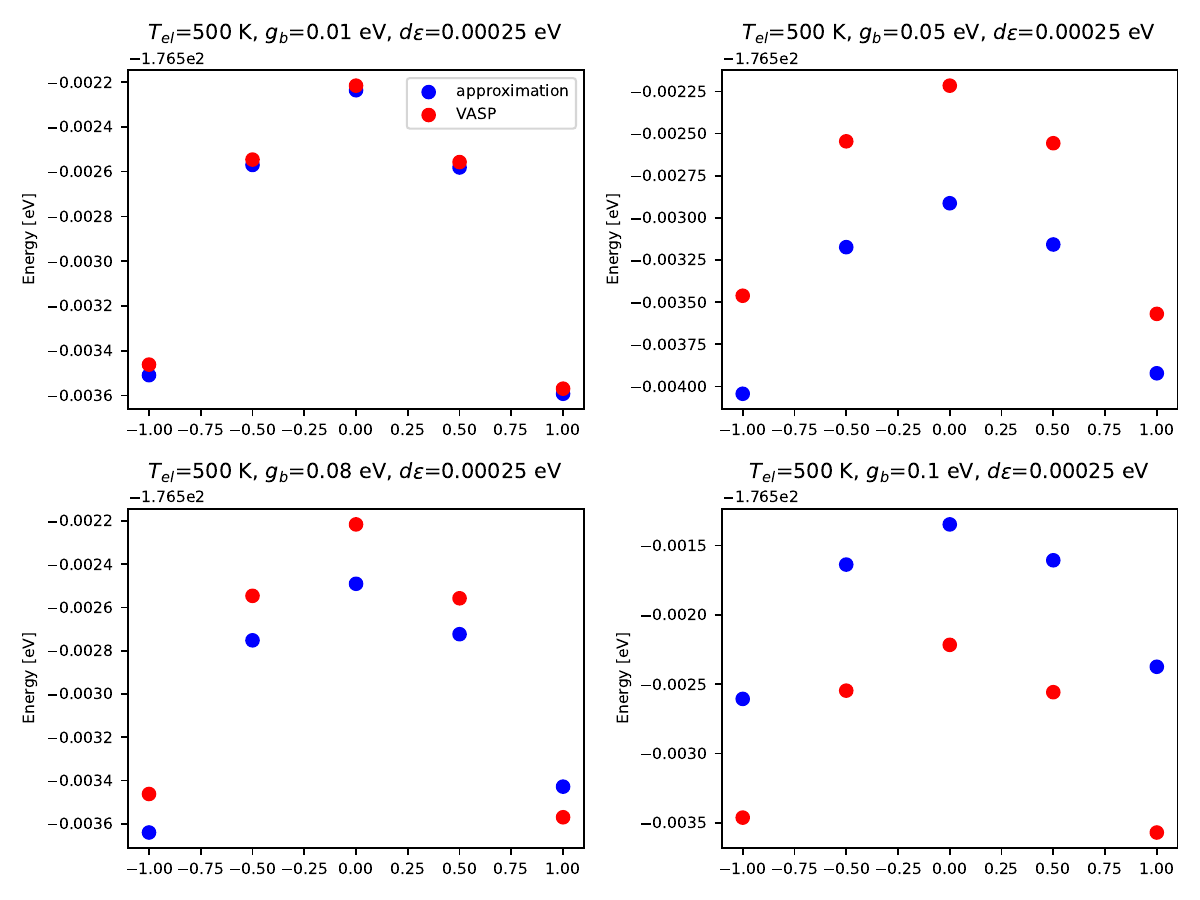}
    \caption{Comparison between the approximation \eqref{eq:S33} and the VASP results for the Helmholtz electronic free energy calculated for one $3\times3\times1$ structure along the $\boldsymbol{q}=2/3~\Gamma$M CDW phonon mode for $T_{el}=500$ K, $d\epsilon=0.00025$ eV and four different $g_{b}$ hyperparameters.}
    \label{fig:diff_g_b}
\end{figure}

Now, we can derive the expression for the force on the $n$-th ion $\boldsymbol{F}_{n}(\vec{\boldsymbol{R}}; T_{el})$ by taking the negative gradient with respect to $\boldsymbol{R}_{n}$ of \eqref{eq:S31}
\begin{align}
    & \hspace{2.25cm} \boldsymbol{F}_{n}(\vec{\boldsymbol{R}}; T_{2}) = -\nabla_{\boldsymbol{R}_{n}}F(\vec{\boldsymbol{R}}; T_{2}) \approx -\nabla_{\boldsymbol{R}_{n}}E^{T_{1}}(\vec{\boldsymbol{R}}; T_{1})\nonumber \\
    & \hspace{0.4cm} - \nabla_{\boldsymbol{R}_{n}}\int\epsilon g^{T_{1}}(\vec{\boldsymbol{R}}; \epsilon)\left[ f\left(\frac{\epsilon - \mu^{T_{1}}(\vec{\boldsymbol{R}}; T_{2})}{k_{B}T_{2}}\right) - f\left(\frac{\epsilon - \mu^{T_{1}}(\vec{\boldsymbol{R}}; T_{1})}{k_{B}T_{1}}\right) \right] d\epsilon \nonumber \\
    & \hspace{2.5cm} - k_{B}T_{2}\nabla_{\boldsymbol{R}_{n}}\int g^{T_{1}}(\vec{\boldsymbol{R}}; \epsilon)s\left(\frac{\epsilon - \mu^{T_{1}}(\vec{\boldsymbol{R}}; T_{2})}{k_{B}T_{2}}\right)d\epsilon~.\label{eq:S34}
\end{align}
We start by calcuating
\begin{align}
    & \nabla_{\boldsymbol{R}_{n}} \int \epsilon g^{T_{1}}(\vec{\boldsymbol{R}}; \epsilon) f\left( \frac{\epsilon-\mu^{T_{1}}(\vec{\boldsymbol{R}}; T_{2})}{k_{B}T_{2}} \right) d\epsilon = \int d\epsilon \epsilon f\left( \frac{\epsilon-\mu^{T_{1}}(\vec{\boldsymbol{R}}; T_{2})}{k_{B}T_{2}} \right)\nabla_{\boldsymbol{R}_{n}}g^{T_{1}}(\vec{\boldsymbol{R}}; \epsilon) \nonumber \\
    & \hspace{2.25cm}+ \int d\epsilon \epsilon g^{T_{1}}(\vec{\boldsymbol{R}}; \epsilon) \frac{\partial f}{\partial\mu^{T_{1}}}\left( \frac{\epsilon-\mu^{T_{1}}(\vec{\boldsymbol{R}}; T_{2})}{k_{B}T_{2}} \right)\nabla_{\boldsymbol{R}_{n}}\mu^{T_{1}}(\vec{\boldsymbol{R}}; T_{2})~,\label{eq:S35}
\end{align}
now we need an expression for $\nabla_{\boldsymbol{R}_{n}}\mu^{T_{1}}(\vec{\boldsymbol{R}}; T_{2})$, which we get by taking the gradient $\nabla_{\boldsymbol{R}_{n}}$ of the expression \eqref{eq:S7}
\begin{align}
    & \hspace{1cm} 0 = \nabla_{\boldsymbol{R}_{n}}\int g^{T_{1}}(\vec{\boldsymbol{R}}; \epsilon)f\left(\frac{\epsilon - \mu^{T_{1}}(\vec{\boldsymbol{R}}; T_{2})}{k_{B}T_{2}}\right)d\epsilon = \int d\epsilon f\left( \frac{\epsilon-\mu^{T_{1}}(\vec{\boldsymbol{R}}; T_{2})}{k_{B}T_{2}} \right)\nabla_{\boldsymbol{R}_{n}}g^{0}(\vec{\boldsymbol{R}}; \epsilon) \nonumber \\
    & \hspace{3.25cm} + \nabla_{\boldsymbol{R}_{n}}\mu^{T_{1}}(\vec{\boldsymbol{R}}; T_{2}) \int d\epsilon g^{T_{1}}(\vec{\boldsymbol{R}}; \epsilon)\frac{\partial f}{\partial\mu^{T_{1}}}\left( \frac{\epsilon-\mu^{0}(\vec{\boldsymbol{R}}; T_{2})}{k_{B}T_{2}} \right) \nonumber \\
    & \hspace{3cm} \implies \nabla_{\boldsymbol{R}_{n}}\mu^{T_{1}}(\vec{\boldsymbol{R}}; T_{2}) = - \frac{\int d\epsilon f\left( \frac{\epsilon-\mu^{0}(\vec{\boldsymbol{R}}; T_{2})}{k_{B}T_{2}} \right)\nabla_{\boldsymbol{R}_{n}}g^{T_{1}}(\vec{\boldsymbol{R}}; \epsilon)}{\int d\epsilon g^{T_{1}}(\vec{\boldsymbol{R}}; \epsilon)\frac{\partial f}{\partial\mu^{T_{1}}}\left( \frac{\epsilon-\mu^{T_{1}}(\vec{\boldsymbol{R}}; T_{2})}{k_{B}T_{2}} \right)}~.\label{eq:S36}
\end{align}
Now, by using \eqref{eq:S36} in \eqref{eq:S35} we have
\begin{align}
    & \nabla_{\boldsymbol{R}_{n}} \int \epsilon g^{T_{1}}(\vec{\boldsymbol{R}}; \epsilon) f\left( \frac{\epsilon-\mu^{T_{1}}(\vec{\boldsymbol{R}}; T_{2})}{k_{B}T_{2}} \right) d\epsilon = \int d\epsilon \epsilon f\left( \frac{\epsilon-\mu^{T_{1}}(\vec{\boldsymbol{R}}; T_{2})}{k_{B}T_{2}} \right)\nabla_{\boldsymbol{R}_{n}}g^{T_{1}}(\vec{\boldsymbol{R}}; \epsilon) \nonumber \\
    & \hspace{0.6cm} + \underbrace{\frac{\int d\epsilon'\epsilon'g^{T_{1}}(\vec{\boldsymbol{R}}; \epsilon')\frac{\partial f}{\partial\mu^{T_{1}}}\left( \frac{\epsilon'-\mu^{T_{1}}(\vec{\boldsymbol{R}}; T_{2})}{k_{B}T_{2}} \right)}{\int d\epsilon'g^{T_{1}}(\vec{\boldsymbol{R}}; \epsilon')\frac{\partial f}{\partial\mu^{T_{1}}}\left( \frac{\epsilon'-\mu^{T_{1}}(\vec{\boldsymbol{R}}; T_{2})}{k_{B}T_{2}} \right)}}_{\equiv\Sigma^{T_{1}}(\vec{\boldsymbol{R}}; T_{2})}\int d\epsilon f\left( \frac{\epsilon-\mu^{T_{1}}(\vec{\boldsymbol{R}}; T_{2})}{k_{B}T_{2}} \right)\nabla_{\boldsymbol{R}_{n}}g^{T_{1}}(\vec{\boldsymbol{R}}; \epsilon) \nonumber \\
    & \hspace{1cm} \implies \nabla_{\boldsymbol{R}_{n}} \int \epsilon g^{T_{1}}(\vec{\boldsymbol{R}}; \epsilon) f\left( \frac{\epsilon-\mu^{T_{1}}(\vec{\boldsymbol{R}}; T_{2})}{k_{B}T_{2}} \right) d\epsilon = \int d\epsilon \left( \epsilon - \Sigma^{T_{1}}(\vec{\boldsymbol{R}}; T_{2}) \right) \nonumber \\
    & \hspace{4.5cm} \times f\left( \frac{\epsilon-\mu^{T_{1}}(\vec{\boldsymbol{R}}; T_{2})}{k_{B}T_{2}} \right)\nabla_{\boldsymbol{R}_{n}}g^{T_{1}}(\vec{\boldsymbol{R}}; \epsilon)~.\label{eq:S37}
\end{align}
In \eqref{eq:S37} we have defined
\begin{align}
    \Sigma^{T_{1}}(\vec{\boldsymbol{R}}; T_{2}) = \frac{\int d\epsilon\epsilon g^{T_{1}}(\vec{\boldsymbol{R}}; \epsilon)\frac{\partial f}{\partial\mu^{T_{1}}}\left( \frac{\epsilon-\mu^{T_{1}}(\vec{\boldsymbol{R}}; T_{2})}{k_{B}T_{2}} \right)}{\int d\epsilon g^{T_{1}}(\vec{\boldsymbol{R}}; \epsilon)\frac{\partial f}{\partial\mu^{T_{1}}}\left( \frac{\epsilon-\mu^{T_{1}}(\vec{\boldsymbol{R}}; T_{2})}{k_{B}T_{2}} \right)}~\label{eq:S38}
\end{align}
%
and we call it the average shift term. In the limit $T_{2}\rightarrow0^{+}$, using \eqref{eq:S4} and the the fact $f^{T_{1}}(0)=\theta(\epsilon^{T_{1}}_{F}(\vec{\boldsymbol{R}})-\epsilon)$, we have
\begin{align}
    \Sigma^{T_{1}}(\vec{\boldsymbol{R}}; 0) = & \frac{\int d\epsilon\epsilon g^{T_{1}}(\vec{\boldsymbol{R}}; \epsilon)\frac{\partial\theta(\epsilon^{T_{1}}_{F}({\vec{\boldsymbol{R}}}) - \epsilon)}{\partial\epsilon^{T_{1}}_{F}({\vec{\boldsymbol{R}}})}}{\int d\epsilon g^{T_{1}}(\vec{\boldsymbol{R}}; \epsilon)\frac{\partial\theta(\epsilon^{T_{1}}_{F}({\vec{\boldsymbol{R}}}) - \epsilon)}{\partial\epsilon^{T_{1}}_{F}({\vec{\boldsymbol{R}}})}} = \frac{\int d\epsilon\epsilon g^{T_{1}}(\vec{\boldsymbol{R}}; \epsilon)\delta(\epsilon^{T_{1}}_{F}(\vec{\boldsymbol{R}})-\epsilon)}{\int d\epsilon g^{T_{1}}(\vec{\boldsymbol{R}}; \epsilon)\delta(\epsilon^{T_{1}}_{F}(\vec{\boldsymbol{R}})-\epsilon)} = \epsilon^{T_{1}}_{F}(\vec{\boldsymbol{R}}) \nonumber \\
    & \hspace{2cm} \implies \Sigma^{T_{1}}(\vec{\boldsymbol{R}}; 0) = \epsilon^{T_{1}}_{F}(\vec{\boldsymbol{R}})~,\label{eq:S39}
\end{align}
which in the limit $T_{1}\rightarrow0^{+}$ becomes
\begin{equation}
    \Sigma^{0}(\vec{\boldsymbol{R}}; 0) = \epsilon^{0}_{F}(\vec{\boldsymbol{R}})~.\label{eq:S40}
\end{equation}
Now, we can calculate the following
\begin{align}
    & \nabla_{\boldsymbol{R}_{n}} \int g^{T_{1}}(\vec{\boldsymbol{R}}; \epsilon)s\left(\frac{\epsilon-\mu^{T_{1}}(\vec{\boldsymbol{R}}; T_{2})}{k_{B}T_{2}}\right) d\epsilon = \int d\epsilon s\left(\frac{\epsilon-\mu^{T_{1}}(\vec{\boldsymbol{R}}; T_{2})}{k_{B}T_{2}}\right) \nabla_{\boldsymbol{R}_{n}}g^{T_{1}}(\vec{\boldsymbol{R}}; \epsilon) \nonumber \\
    & \hspace{2.25cm}+ \int d\epsilon g^{T_{1}}(\vec{\boldsymbol{R}}; \epsilon)\frac{\partial s}{\partial\mu^{T_{1}}}\left(\frac{\epsilon-\mu^{T_{1}}(\vec{\boldsymbol{R}}; T_{2})}{k_{B}T_{2}}\right)\nabla_{\boldsymbol{R}_{n}}\mu^{T_{1}}(\vec{\boldsymbol{R}}; T_{el})~.\label{eq:S41}
\end{align}
Furthermore. we have
\begin{align}
    & \hspace{3cm} \frac{\partial s}{\partial\mu^{T_{1}}}\left(\frac{\epsilon-\mu^{T_{1}}(\vec{\boldsymbol{R}}; T_{2})}{k_{B}T_{2}}\right) = \frac{\partial s}{\partial f}\frac{\partial f}{\partial \mu^{T_{1}}}\left(\frac{\epsilon-\mu^{T_{1}}(\vec{\boldsymbol{R}}; T_{2})}{k_{B}T_{2}}\right) \nonumber \\
    & \hspace{-0.25cm} \stackrel{\eqref{eq:S32}}{=} \left(\ln\left(\frac{f}{1-f}\right)\right)\frac{\partial f}{\partial\mu^{T_{1}}}\left(\frac{\epsilon-\mu^{T_{1}}(\vec{\boldsymbol{R}}; T_{2})}{k_{B}T_{2}}\right) \stackrel{\eqref{eq:S2}}{=} - \frac{\epsilon - \mu^{T_{1}}(\vec{\boldsymbol{R}}; T_{2})}{k_{B}T_{2}}\frac{\partial f}{\partial\mu^{T_{1}}}\left(\frac{\epsilon-\mu^{T_{1}}(\vec{\boldsymbol{R}}; T_{2})}{k_{B}T_{2}}\right)~.\label{eq:S42}
\end{align}
Now, by inserting \eqref{eq:S42} into \eqref{eq:S41}, and using \eqref{eq:S36} and \eqref{eq:S38} we have
\begin{align}
    & \nabla_{\boldsymbol{R}_{n}} \int g^{T_{1}}(\vec{\boldsymbol{R}}; \epsilon)s\left(\frac{\epsilon-\mu^{T_{1}}(\vec{\boldsymbol{R}}; T_{2})}{k_{B}T_{2}}\right) d\epsilon = \int \nabla_{\boldsymbol{R}_{n}}g^{T_{1}}(\vec{\boldsymbol{R}}; \epsilon)\left[ s\left(\frac{\epsilon-\mu^{T_{1}}(\vec{\boldsymbol{R}}; T_{2})}{k_{B}T_{2}}\right) \right. \nonumber \\
    & \hspace{2.75cm} + \left. \frac{\Sigma^{T_{1}}(\vec{\boldsymbol{R}}; T_{2}) - \mu^{T_{1}}(\vec{\boldsymbol{R}}; T_{2})}{k_{B}T_{2}}f\left( \frac{\epsilon-\mu^{T_{1}}(\vec{\boldsymbol{R}}; T_{2})}{k_{B}T_{2}} \right)\right] d\epsilon~.\label{eq:S43}
\end{align}
Finally, by using \eqref{eq:S34}, \eqref{eq:S37} and \eqref{eq:S43} we arrive at
\begin{align}
    & \hspace{0.25cm} \boldsymbol{F}_{n}(\vec{\boldsymbol{R}}; T_{2}) \approx -\nabla_{\boldsymbol{R}_{n}}E^{T_{1}}(\vec{\boldsymbol{R}}; T_{1}) -\int d\epsilon \nabla_{\boldsymbol{R}_{n}}g^{T_{1}}(\vec{\boldsymbol{R}}; \epsilon)\left[\left( \epsilon - \Sigma^{T_{1}}(\vec{\boldsymbol{R}}; T_{2}) \right)f\left( \frac{\epsilon-\mu^{T_{1}}(\vec{\boldsymbol{R}}; T_{2})}{k_{B}T_{2}} \right) \right. \nonumber \\
    & \left. \hspace{0.25cm} - \left( \epsilon - \Sigma^{T_{1}}(\vec{\boldsymbol{R}}; T_{1}) \right)f\left( \frac{\epsilon-\mu^{T_{1}}(\vec{\boldsymbol{R}}; T_{1})}{k_{B}T_{1}} \right)\right] - k_{B}T_{2}\int d\epsilon \nabla_{\boldsymbol{R}_{n}}g^{T_{1}}(\vec{\boldsymbol{R}}; \epsilon)\left[s\left(\frac{\epsilon-\mu^{T_{1}}(\vec{\boldsymbol{R}}; T_{2})}{k_{B}T_{2}}\right) \right. \nonumber \\
    & \left. \hspace{4cm} + \frac{\Sigma^{T_{1}}(\vec{\boldsymbol{R}}; T_{2}) - \mu^{T_{1}}(\vec{\boldsymbol{R}}; T_{2})}{k_{B}T_{2}}f\left( \frac{\epsilon-\mu^{T_{1}}(\vec{\boldsymbol{R}}; T_{2})}{k_{B}T_{2}} \right)\right]~.\label{eq:S44}
\end{align}
Or, using $T_{1}\rightarrow0^{+}$ and the definitions from \eqref{eq:S27}, along with \eqref{eq:S4} and \eqref{eq:S40}, we have
\begin{align}
    & \hspace{0.25cm} \boldsymbol{F}_{n}(\vec{\boldsymbol{R}}; T_{el}) \approx -\nabla_{\boldsymbol{R}_{n}}E^{0}(\vec{\boldsymbol{R}}; 0) -\int d\epsilon \nabla_{\boldsymbol{R}_{n}}g^{0}(\vec{\boldsymbol{R}}; \epsilon)\Biggl[\left( \epsilon - \Sigma^{0}(\vec{\boldsymbol{R}}; T_{el}) \right)f\left( \frac{\epsilon-\mu^{0}(\vec{\boldsymbol{R}}; T_{el})}{k_{B}T_{el}} \right) \nonumber \\
    & \hspace{1cm} - \left( \epsilon - \epsilon^{0}_{F}(\vec{\boldsymbol{R}})\right)\theta(\epsilon^{0}_{F}(\vec{\boldsymbol{R}})-\epsilon)\Biggr] - k_{B}T_{el}\int d\epsilon \nabla_{\boldsymbol{R}_{n}}g^{0}(\vec{\boldsymbol{R}}; \epsilon) \left[ s\left(\frac{\epsilon-\mu^{0}(\vec{\boldsymbol{R}}; T_{el})}{k_{B}T_{el}}\right) \right. \nonumber \\
    & \left. \hspace{3.75cm} + \frac{\Sigma^{0}(\vec{\boldsymbol{R}}; T_{el}) - \mu^{0}(\vec{\boldsymbol{R}}; T_{el})}{k_{B}T_{el}}f\left( \frac{\epsilon-\mu^{0}(\vec{\boldsymbol{R}}; T_{el})}{k_{B}T_{el}} \right) \right]~,\label{eq:S45}
\end{align}
where $f^{0}(0)=\theta(\epsilon^{0}_{F}(\vec{\boldsymbol{R}})-\epsilon)$ was used.
\clearpage
In Fig. \ref{fig:subplots_approximation}, we show the comparison between the approximation \eqref{eq:S33} and the VASP results calculated along the $\boldsymbol{q}=2/3~\Gamma$M CDW phonon mode in $3\times3\times1$ supercell.
\clearpage
In Fig. \ref{fig:diff_d_eps} we report the comparison between the approximation \eqref{eq:S33} and the VASP results for the Helmholtz electronic free energy calculated for one $3\times3\times1$ structure along the $\boldsymbol{q}=2/3~\Gamma$M CDW phonon mode for $T_{el}=500$ K, $g_{b}=0.01$ eV and four different $d\epsilon$ hyperparameters. As can be seen, smaller values of $d\epsilon$ result in better agreement between the approximation and the VASP results. 
\clearpage
In Fig. \ref{fig:diff_g_b} we report the comparison between the approximation \eqref{eq:S33} and the VASP results for the Helmholtz electronic free energy calculated for one $3\times3\times1$ structure along the $\boldsymbol{q}=2/3~\Gamma$M CDW phonon mode for $T_{el}=500$ K, $d\epsilon=0.00025$ eV and four different $g_{b}$ hyperparameters. As can be seen, smaller values of $g_{b}$ result in better agreement between the approximation and the VASP results.

\clearpage
\section{SOAP and KRR hyperparameter optimization}
\label{S_soap_krr}
The ML model for the finite $T_{el}$ part of the Helmholtz electronic free energy given by \eqref{eq:S33} depends on both SOAP and KRR hyperparameters. We have performed the hyperparameter optimization of both of them at once using Optuna software \cite{akiba2019optunanextgenerationhyperparameteroptimization}.
We have used TPESampler \cite{tpesampler}, which is a Bayesian algorithm for hyperparameter optimization. The optimization was done in $100$ steps, where in every step the training set was randomly split into temporary training and validation sets in the ratio 80\% to 20\%. This splitting was done using a different random seed in each step. We have kept test set comprised of 20\% of the original data untouched until the last inference on it to get the unbiased metrics about the models' performance. The objective function during the optimization was given by Eq.~\eqref{eq:rmse} in the main text and the goal was to minimize this objective function evaluated on the validation set.

Hyperparameters we wanted to optimize were both SOAP and KRR ones, because this system to our knowledge has not been treated within this workflow, so we could not find SOAP hyperparameters that we could use as a starting point.

For constructing descriptors we have chosen the power spectrum SOAP representation, within it we had to optimize the cutoff radius (7.6 \AA), maximum radial (10) and maximum angular (10) numbers \cite{PhysRevB.87.184115}. Furthermore, for representing ionic environments we have used Gaussian type orbitals of constant width, which also was a part of optimization process (0.7 \AA). To optimize the hyperparameters that control the smoothness of decay of the interactions we have used radial scaling option \cite{C8CP05921G} and we optimized hyperparameters that come with it (scale$=6.84$ \AA, rate=$1.2$, exponent$=1$).

Regarding the KRR hyperparameters, we had to optimize the number of sparse ionic environments out of total number of ions for each ionic species in our system. Since it would take very long time to train models if we had taken these numbers of sparse environments too big, we have chosen to optimize them in range from 0.5\% to 1.25\% of total number of ions for each ionic species (Se: 545 ionic environments, Nb: 272 ionic environments). Furthermore, there is a parameter that effectively includes higher than two-body interactions even when we use power spectrum, namely to construct the kernels we take the dot product between SOAP vectors and raise it to some power, called zeta ($\zeta=2$). The equation for weights in KRR model is given by Eq. (11) in \cite{acs.chemrev.1c00022}. A way to solve this equation in numerically stable way, as implemented in \textit{librascal}, is to choose the most relevant eigenvalues of the $K_{\vec{A}~\vec{A}}$ matrix from the main text. This is controlled by threshold parameter ($5.62\times10^{-12}$). This method is similar to principal component analysis.

Above we have written in parentheses the values of all hyperparameters of the "ML used" model from the main text. 
\begin{figure}[!h]
\centering
\includegraphics[width=\columnwidth]{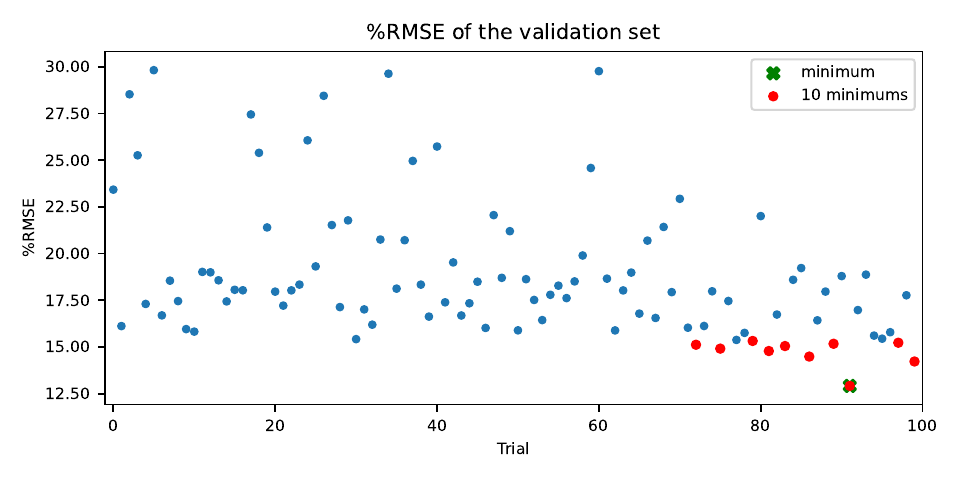}
\caption{Distribution of $\%$RMSE evaluated on the validation set during the hyperparamter optimization process.}
\label{fig:optuna}
\end{figure}
After the minimization of the $\%$RMSE on the validation set, the models were trained again using the total training set and at the end evaluated on the test set. 

In Fig. \ref{fig:optuna} we report the distribution of $\%$RMSE on the validation set during the hyperparameter optimization process. Also, the 10 minimal errors are shown in red (in green we show the model with minimum \%RMSE), those models were used to make an ensemble explained in the main text.
%
%
%
\clearpage
\begin{figure}[!h]
\centering
\includegraphics[scale=1.25]{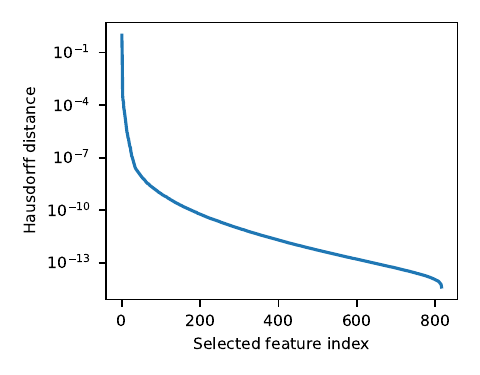}
\caption{Plot of the Hausdorff distance used for selecting the number of relevant eigenvalues of the $K_{\vec{A}~\vec{A}}$ matrix for the "ML used" model.}
\label{fig:hausdorff}
\end{figure}
\begin{figure}[!h]
\centering
\includegraphics[scale=1.25]{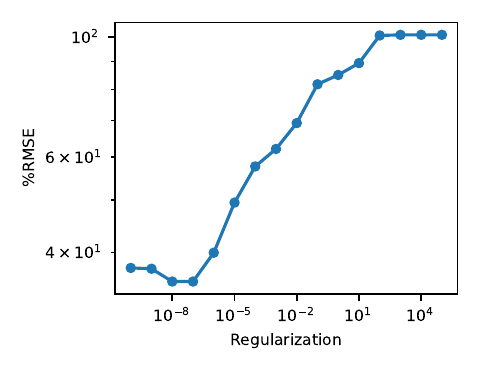}
\caption{Plot of the different regularization parameters and their \%RMSE for the "ML used" model.}
\label{fig:reg1}
\end{figure}
\clearpage
In Figs. \ref{fig:hausdorff} we report the plot of the Hausdorff distance for "ML used" model used for selecting the number of relevant eigenvalues of the $K_{\vec{A}~\vec{A}}$ matrix. Earlier explained threshold parameter is the Hausdorff distance chosen from this plot.

In Fig. \ref{fig:reg1} we show \%RMSE vs. regularization plot, that helps us to determine the L2 type regularization parameter from the KRR model. This plot was obtained by 5-fold cross-validation and the minimum of this curve is taken as the best regularization parameter for a given model. The plot given here also corresponds to "ML used" model and the minimum of the curve is at $10^{-7}$.

In literature one can also find a method \cite{natarajan2020particleswarmbasedhyperparameter} where SOAP and KRR hyperpearameters are optimized in separate phases. We have not used this method in this work.

\clearpage
\section{Phonon dispersions in the harmonic approximation}
\label{S_harm_approx}
\begin{figure}[!h]
\centering
\includegraphics[scale=1.25]{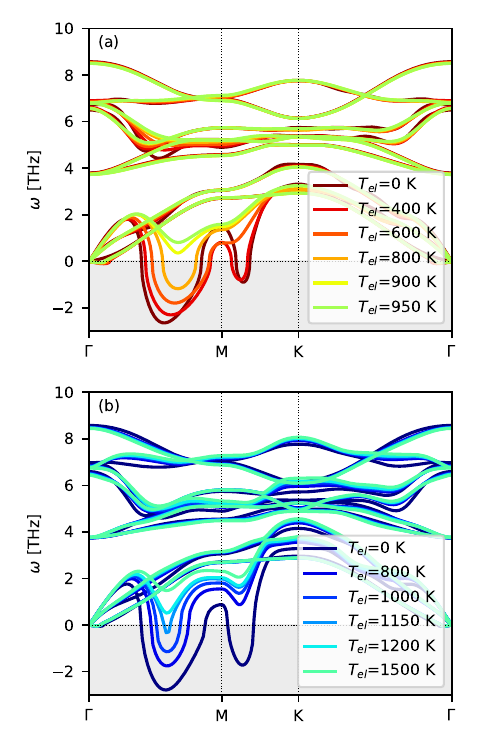}
\caption{(a) All 9 phonon dispersions calculated using the VASP. (b) All 9 phonon dispersions calculated using ML model from the main text.}
\label{fig:subplots_harmonic_full}
\end{figure}

In Fig. \ref{fig:subplots_harmonic_full} we report all of the nine phonon dispersions (three acoustic and six optical branches) calculated in the harmonic approximation with both the VASP and the full interatomic ML potential from the main text. One can see that optical phonon dispersions are even in a better agreement between VASP and ML calculations.

\section{Two temperature model with DFT input parameters}

The energy flow between electron and phonon subsystems, represented with the corresponding temperatures $T_{el}=T_{el}(t)$ and $T_{ion}=T_{ion}(t)$, can be simulated using the following coupled equations\,\cite{caruso22,novko20}
\begin{eqnarray}
\frac{\partial  T_{el}}{\partial t} &=& \frac{I(t)}{C_{el}(T_{el})}- \frac{G}{C_{el}(T_{el})}(T_{el}-T_{ion})~,
\label{eq:eq1}
\\
\frac{\partial  T_{ion}}{\partial t} &=& \frac{G}{C_{ion}(T_{ion})}(T_{el}-T_{ion})~.
\label{eq:eq2}
\end{eqnarray}
The corresponding specific heats 
$C_{el}(T_{el})$ and $C_{ion}(T_{ion})$ are 
defined as: 
\begin{eqnarray}
C_{el}(T_{el})
&=&
\int_{-\infty}^{\infty}
d\varepsilon N(\varepsilon)\varepsilon
\frac{\partial f(\varepsilon;T_{el})}{\partial T_{el}}~,
\label{eq:eq4}
\\
C_{ion}(T_{ion})
&=&
\int_{0}^{\infty}
d\omega F(\omega)\omega
\frac{\partial n(\omega;T_{ion})}{\partial T_{ion}}~,
\label{eq:eq5}
\end{eqnarray}
where $N(\varepsilon)$ is electron and $F(\omega)$ is phonon density of states. The electron-phonon relaxation rate $G$ is calculated from density functional perturbation theory (DFPT) and  electron-phonon calculations as 
\begin{align}
G=\frac{\pi k_B}{\hbar N(\varepsilon_F)}\lambda\left\langle \omega^2 \right\rangle\int_{-\infty}^{\infty} d\varepsilon N^2(\varepsilon)\left(- \frac{\partial f(\varepsilon;T_{el})}{\partial \varepsilon} \right)~,
\label{eq:eq7}
\end{align}
where $\lambda$ is the total electron-phonon coupling strength,
\begin{align}
\lambda=2\int d\Omega \frac{\alpha^2 F (\Omega)}{\Omega}~,
\label{eq:eq7_0}
\end{align}
and $\left\langle \omega^2 \right\rangle$
is the corresponding second moments of the phonon spectrum, 
\begin{align}
\left\langle \omega^2 \right\rangle=\frac{2}{\lambda}\int d\Omega \Omega\alpha^2 F (\Omega)~.
\label{eq:eq7_1}
\end{align}
Here $\alpha^2 F(\Omega)$ is the Eliashberg function, which represents the strength of electron-phonon coupling for each phonon energy $\Omega$.

\begin{figure}[!h]
\centering
\includegraphics[width=0.7\textwidth]{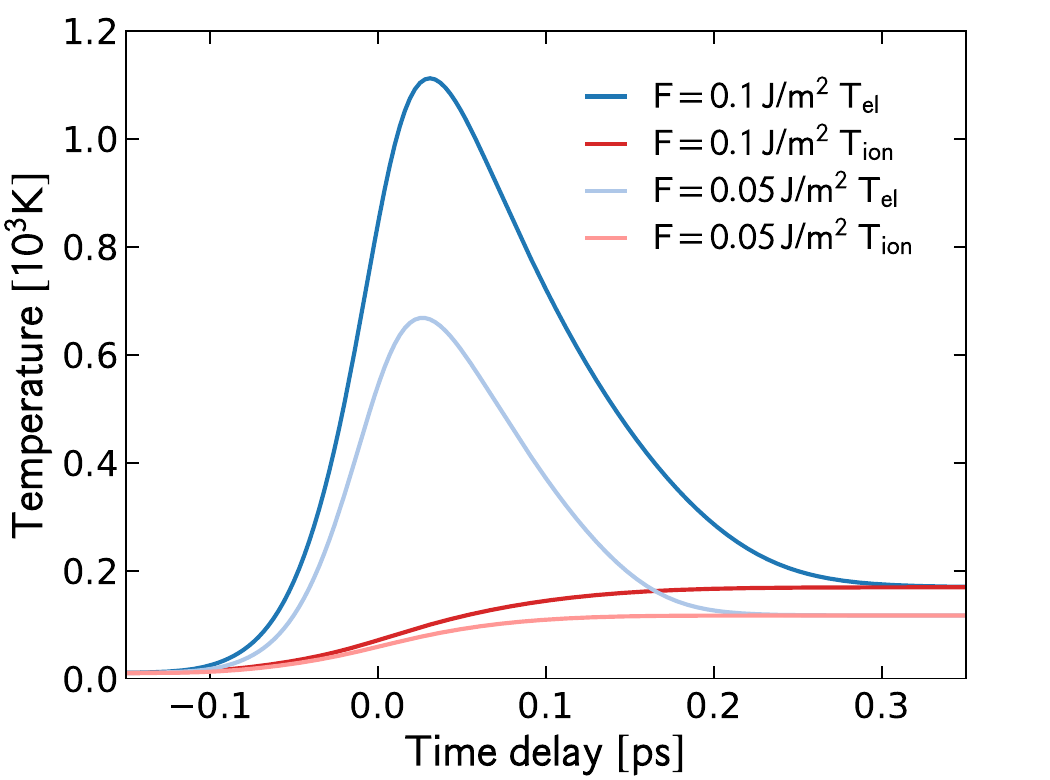}
\caption{The results of the TTM calculations in NbSe$_2$ when two different fluences are applied.}
\label{fig:sm_ttm2}
\end{figure}

In addition, $I(t)$ describes a femtosecond Gaussian pump pulse with fluence $F$ and duration $t_p$,
\begin{align}
I(t)=\frac{2F}{t_p}\sqrt{\frac{\log{2}}{\pi}}\exp{\left[ -4\log{2}\left( \frac{t}{t_p} \right)^2 \right]}~.
\label{eq:eq7_2}
\end{align}
All the input parameters needed for the above TTM equations (except $F$ and $t_p$) are obtained from DFT and DFPT calculations as implemented in \textsc{Quantum ESPRESSO} (QE) package~\cite{Giannozzi2017}. The electron-phonon coupling constants were obtained with Wannier interpolation of EPC matrix elements as implemented in the EPW code~\cite{Marzari2012, Noffsinger2010, Ponce2016}. In order to reduce the differences between the two type of calculations, here we chose the same structural parameters for NbSe$_2$ as the one used in VASP calculations of ML interatomic potential. For the same reason, we used the PBE functional \cite{PhysRevLett.77.3865}. The DFPT calculations of the phonons were done on coarse grids of $\mathbf{k}=24\times24\times1$ and $\mathbf{q}=6\times6\times1$. We used Fermi-Dirac smearing with a broadening of 1600\,K. For the electron-phonon calculations within EPW, we use 11 maximally localized Wannier functions\,\cite{Marzari2012} with the initial projections of d orbitals on Nb and p orbitals on the Se atoms. To calculate Eliashberg function $\alpha^2 F(\omega)$ we use a fine grids of $\mathbf{k}=240\times240\times1$ and $\mathbf{q}=80\times80\times1$.
\begin{figure}[!h]
\centering
\includegraphics[width=0.7\textwidth]{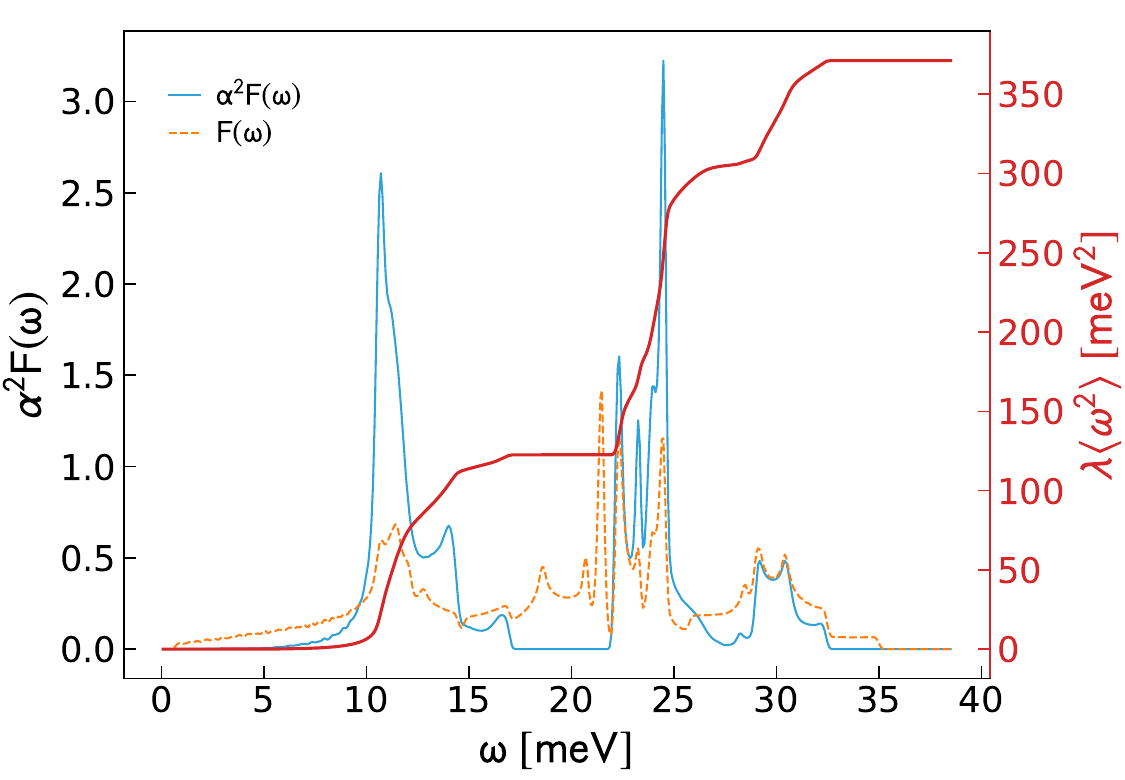}
\caption{Theoretical results for Eliashberg function $\alpha^2 F(\omega)$ (blue full line) and phonon density of states $F(\omega)$ (orange dashed line) in NbSe$_2$, with the corresponding value for the second moments of the phonon spectrum $\left\langle \omega^2 \right\rangle$ (red).
}
\label{fig:sm_ttm1}
\end{figure}
In Fig.\,\ref{fig:sm_ttm1} we show the results for $\alpha^2 F(\omega)$ and phonon density of states $F(\omega)$. From this we get $\lambda=1.3$ and $\lambda\langle\omega^2\rangle=371$\,meV. The obtained value of $\lambda$ is in a good agreement with estimation from the ARPES measurements, where the momentum-dependent $\lambda$ is between 0.9 and 1.9\,\cite{valla2004}, as well as with the previous DFT-based calculation\,\cite{anikin2020}. Further, in Fig.\,\ref{fig:sm_ttm2} we show the time evolution of electron and phonon temperatures as obtained with TTM and the DFT-derived input parameters for two different values of fluence $F$. For pulse width we took $t_p=30$\,fs, which is very close to the experimental width\,\cite{anikin2020,venturini2023}. For $F=0.05$\,J/m$^2$ we get the maximal values of $T_{el}\approx600$\,K and $T_{ion}\approx100$\,K, while for $F=0.1$\,J/m$^2$ we get the maximal values of $T_{el}\approx1000$\,K and $T_{ion}\approx170$\,K. Since we obtain that the CDW phase is melted above $T_{ion}\approx100$\,K and $T_{el}=800$\,K, we consider that the critical value of fluence for melting the CDW phase in NbSe$_2$ is $F=0.05-0.1$\,J/m$^2$, which agrees well with the pump-probe optical spectroscopy measurements\,\cite{venturini2023}.

\clearpage
\bibliography{ref}